\documentclass[manuscript]{aastex}
\usepackage{url}
\usepackage{natbib}
\usepackage{hyperref}
\usepackage{journal}

\begin{document}

\title{The Horizontal Branch population of NGC 1851 as revealed by the Ultra-violet Imaging Telescope (UVIT)}

\author{Annapurni Subramaniam$^{1}$, 
Snehalata Sahu$^{1}$,
Joseph E. Postma$^{2}$,
Patrick C\^ot\'e $^{3}$,
J.B.~Hutchings$^{3}$,
N. Darukhanawalla$^{3}$,
Chung Chul$^{4}$,
S.N. Tandon$^{1,5}$, 
N. Kameswara Rao$^{1}$,
K. George$^{1}$, 
S.K. Ghosh$^{6,7}$, 
V. Girish$^{8}$, 
R. Mohan$^{1}$,
J. Murthy$^{1}$, 
A.K. Pati$^{1}$,
K. Sankarasubramanian$^{1,8,9}$,
C.S. Stalin$^{1}$ \and
S. Choudhury$^{10}$
}
\affil{$^1$Indian Institute of Astrophysics, Koramangala II Block, Bangalore-560034, India\\
$^2$University of Calgary, 2500 University Drive NW, Calgary, Alberta Canada\\
$^3$National Research Council of Canada, Herzberg Astronomy and Astrophysics Program, 5071 West Saanich Road, Victoria, BC, V9E 2E7, Canada\\
$^4$ Center for Galaxy Evolution Research, Yonsei University, Seoul 03722, Korea \\
$^5$Inter-University Centre for Astronomy and Astrophysics, Pune, India\\
$^6$National Centre for Radio Astrophysics, Pune, India\\
$^7$Tata Institute of Fundamental Research, Mumbai, India\\
$^8$ISRO Satellite Centre, HAL Airport Road, Bangalore 560017\\
$^9$Center of Excellence in Space Sciences India, Indian Institute of Science Education and Research (IISER), Kolkata,
Mohanpur 741246, West Bengal, India\\
$^{10}$ Yonsei University Observatory, 120-749, Seoul, Republic of Korea \\}
\email{$^1$purni@iiap.res.in}

\begin{abstract}
We present the UV photometry of the globular cluster NGC 1851 using 
images acquired with the Ultra-violet Imaging Telescope (UVIT) onboard the ASTROSAT satellite.
PSF-fitting photometric data derived from images in two far-UV (FUV) filters and one near-UV (NUV) filter
are used to construct color-magnitude diagrams (CMD), in combination with HST and ground-based optical 
photometry.  In the FUV, we detect only the bluest part of the cluster horizontal branch (HB); in the NUV, we detect 
the full extent of the HB, including the red HB, blue HB  and a small number of RR Lyrae stars. 
UV variability was detected in 18 RR Lyrae stars, and 3 new variables were also detected in the central region.  
The UV/optical CMDs are then compared with 
isochrones of different age and metallicity (generated using Padova and BaSTI models) and synthetic HB (using helium enhanced $Y^2$ models).
We are able to identify two populations among the HB stars, which are found to have  either an age range of 10-12~Gyr, or a range in Y$_{ini}$ of 0.23 - 0.28, for a metallicity of
[Fe/H] =$-$1.2 to $-$1.3. These
estimations from the UV CMDs are consistent with those from optical studies. The almost complete sample of the HB stars tend to show a marginal difference in spatial/azimuthal
distribution among the blue and red HB stars. This study thus show cases the capability of UVIT, with its excellent resolution and large field of view, to study the hot stellar population 
in Galactic globular clusters.
\end{abstract}


\section{Introduction} \label{sec:intro}
Globular clusters (GCs) were long considered to be simple stellar populations, composed of stars having nearly identical age and chemical
composition. However, a succession of studies carried out during the past decade have established that many clusters deviate from these 
assumptions (see, e.g., \cite{Piotto2009, Piotto2012}). NGC 1851 is one such object. Several photometric and spectroscopic 
studies have concluded that this cluster contains multiple stellar populations (e.g., \cite{Milone2008, Carretta2011,
Gratton2012}). The cluster is also remarkable in that it appears to be surrounded by a diffuse stellar halo extending
to radii of $\sim$250~pc \citep{Olszewski2009}. 

An early study of the cluster by \cite{Walker1992} found its core, although unresolved, to have an unusually blue color. At the 
same time, the horizontal branch (HB) was found to be clearly bimodal, with both a red clump and an extended blue tail.
Subsequent HST/ACS photometry of the cluster from \citep{Milone2008} revealed the existence of two distinct 
sub-giant branches (SGBs) as well.  These authors found that the bright SGB component to outnumber the
the fainter component by $\sim$10\%, and that the percentages of red and blue HB stars are 63\% (RHB) and 37\% (BHB), respectively.
These multiple SGB populations were interpreted as evidence either for two populations having either an age difference of about 1--2 Gyr 
or a significant difference in their C+N+O content \citep{Piotto2012, Cassisi2008}.
\cite{Salaris2008}, on the basis of detailed 
synthetic modeling of HB stars, concluded that these two populations are distributed in different regions of the HB. 
The evolved stars, belonging to the bright SGB component, are confined to the red portion of HB whereas 
the the faint SGB population have HB stars distributed from the blue to the red, even 
populating the RR Lyrae instability strip. \cite{Salaris2008} also argued that large variations in the initial 
He abundance between the two sub-populations could be ruled out. 

\cite{Gratton2012} studied the chemical composition and Na-O anti-correlation of HB stars from moderately high 
resolution spectra for 91 stars on the bimodal HB and 13 stars on the red giant Branch (RGB). This spectroscopic
study found that the RHB stars divide into two groups, with the vast majority of stars being Na-poor and O-rich. About 
10 -- 15\% of the stars are Na-rich and moderately O-poor, with most, but not all, being Ba-rich. The two groups were 
found to occupy distinct regions of the CMD, where the Na-rich stars being redder and slightly brighter than the Na-poor 
ones. They found the BHB stars to be enriched in N, but not exceptionally so, and the total CNO abundance 
unlikely to be anomalous.
The He abundance of the BHB was found to be consistent with both the cosmological value and a small
He enhancement. This confirmed the lack of evidence for very large He enhancements within NGC 1851.
Their study identified an age spread of $\sim$ 1.5 Gyr as the only viable explanation for the split of the SGB
and suggested that most of the BHB stars are older than their RHB counterparts 
based on their computed synthetic HB and adopting the bright and faint SGB fractions of \cite{Milone2008}. 

This conclusion is in agreement with the findings of \cite{Carretta2011}, who first suggested that the
cluster might be the product of a merger between two GCs having an age difference of $\sim$1 Gyr. \cite{Han2009} modelled the
HB stars and suggested that the presence of bimodal horizontal-branch distribution in NGC 1851 can be naturally 
reproduced if the metal-rich second generation stars are also enhanced in helium. \cite{JooLee2013} later 
argued that the two populations were similar in age and metallicity, but differed in Helium content, after comparing their
(Yonsei-Yale) models with the observed CMD. \cite{Kunder2013} modelled the HB stars using He enhanced models,
and noted synthetic BHB stars, along with the RR Lyrae stars, having an increasing He content and ages $\sim$ 1.5 Gyr
older than their RHB counterparts. They too argued that the cluster may be a merger remnant. Using deep HST imaging, 
\cite{Piotto2015} presented NUV-optical CMDs for a number of Galactic globular clusters. The CMD of NGC 1851, which is 
shown in their Figure 5, reveals  a broad and complex HB. \cite{Yong2008} also suggested that two stellar populations 
might exist in NGC 1851 and noted that this cluster has experienced a 
complicated formation history bearing some similarities to Omega Centauri. In summary, there is a general agreement that the HB population 
in this cluster shows an age difference, and small differences in [Fe/H], CNO and He enhancement. Among the above parameters, particularly, the differences in age and He content are expected to produce similar effects in the HB morphology. Hence, it has not been possible to have a unique interpretation of HB morphology and population difference. It is also expected that the UV studies help in lifting this degeneracy.

The distribution of HB stars in the UV CMD can be used
to better understand the formation and evolution of these stars (see, e.g., Piotto et al. 2015 and references therein).
A pioneering UV study of 44 GCs using the GALEX images
was presented by \cite{Schiavon2012} who characterized the principal features of the UV CMDs,  established the 
locus of post-core-He burning stars in the UV CMD, and presented a catalog of candidate asymptotic
giant branch (AGB), and post-AGB stars in their sample clusters. In their Figure 11,
\cite{Schiavon2012} presented a FUV vs (FUV$-$NUV) CMD for NGC 1851 that contained a few dozen HB stars and
a handful of candidate hot stars. However, due to crowding issues, this study was restricted to the region beyond 
the central 50\arcsec~and did not include a detailed comparison to stellar evolutionary models.
In this study, we present the results of a pilot study to showcase the capability of Ultraviolet Imaging Telescope (UVIT) 
to study Galactic globular clusters, by
presenting the UV images, photometry which can probe the inner regions of the cluster and the  UV CMDs of NGC 1851.

The paper is arranged as follows. In \S\ref{sec:data} we describe the acquisition, reduction and analysis of 
the UVIT images on which our study is based, including a comparison to earlier GALEX observations. 
In \S\ref{sec:cmd}, we discuss the UV and optical CMDs of NGC 1851 based on our UVIT imaging, as well
as HST and ground-based optical photometry. In \S\ref{sec:analysis}, we characterize the HB
populations in the cluster by comparing our UV/optical CMDs to those generated with two widely used
stellar evolutionary models. In \S\ref{sec:Herich} we investigate the possibility of He enrichment among the BHB stars
using He enhanced models and FUV CMD. In \S\ref{sec:spatial}, we explore the spatial distribution of the distinct HB
populations, and in \S\ref{sec:rr}, we present the first UV light curves for RR Lyrae stars in this cluster.
We conclude in \S\ref{sec:conclusions}.

\section{UVIT Data}
\label{sec:data}

The images used in this study were acquired with the UVIT instrument aboard the ASTROSAT satellite.
UVIT consists of twin 38-cm telescopes --- one for the FUV region (130--180~nm) and the other for
the NUV (200--300~nm) and visible (VIS) regions (320--550~nm). These wavelength ranges
are divided using a dichroic mirror for beam-splitting. UVIT is primarily an imaging instrument, simultaneously 
generating images in the FUV, NUV and VIS channels over a circular 
field of diameter 28$^\prime$.  Each channel can be further divided using a selectable set of filters. Full details 
on the telescope and instruments, including initial calibration results, can be found in \cite{Tandon2016} and 
\cite{Subramaniam2016}. The primary photometric calibration for all FUV and NUV filters 
were performed using observations of HZ4, a white dwarf spectrophotometric standard star \citep{Tandon2017}.
The sensitivity variations across the field are calibrated to produce reliably flattened images that are
accurate to within 3\%, in the case of the FUV filters. The photometric accuracy is thus within 3\% for FUV and within 10\% in 
the inner regions for the NUV, due to flat fielding errors and, based on repeated standard star observations, photometry for individual stars is accurate up to a few percent.

NGC 1851 was observed as part of UVIT's Performance Verification (PV) phase during 19-21 March 2016. The 
observations were part of the multi-wavelength demonstration that focused on NGC 1851 and its hot or
exotic stellar content. 
In this study, we analyze only data from the UVIT instrument.  
In all, three filters were used in the program: F148W and F169M (for the FUV channel) and N279N (for the NUV).
Images were corrected for geometric distortion, flat field illumination and spacecraft drift using our customized 
software package CCDLAB \citep{Postma2017}. In Figure~1, we compare the final FUV image from UVIT 
to that obtained with GALEX; the equivalent comparison for the NUV  is shown in Figure~2.
Isolated stellar sources in the UVIT images have FWHM~$\sim$1.5\arcsec~and $\sim$1.2\arcsec~in the FUV and NUV channels, respectively.
In terms of angular resolution, the UVIT images are thus far superior to those from GALEX (4\farcs5--5\farcs5). In the FUV, 
we are able to resolve stars well into the core of the cluster (although the NUV image still suffers from crowding in the central regions). 
 The combined FUV+UNV false color image of NGC 1851 is presented in Figure~3. This combined image demonstrates the capability
to produce excellent resolution in both the UV channels, including satisfactory correction to satellite drift, distortion and alignment. 
Finally, because the UVIT data were collected over several orbits, we are able to detect a number
of variable stars in the field. In \S\ref{sec:rr}, we discuss these variables and characterise their flux 
variations in the UV region. 

\begin{figure*}[h]
\begin{center}
\includegraphics[scale=4.1]{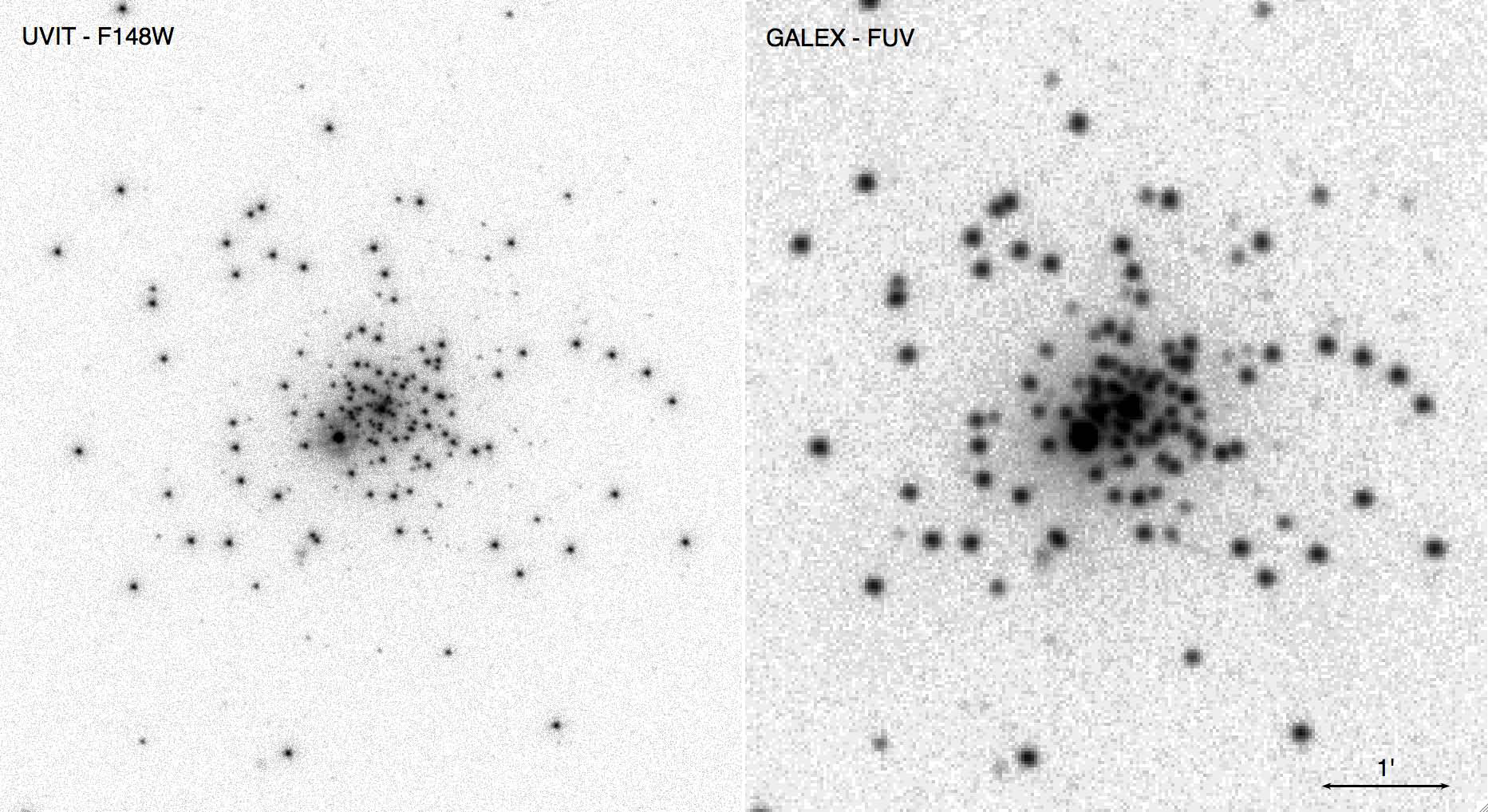}
\caption{Comparison of FUV images for NGC 1851 taken with UVIT and GALEX (left and right panels, respectively).
The UVIT image was constructed from a total integration time of 6982.13 sec in the F148W filter. North is up and east is to
the left. Only a subset of the full image is shown in each case.
}
\end{center}
\end{figure*}
\begin{figure*}[h]
\begin{center}
\includegraphics[scale=4.1]{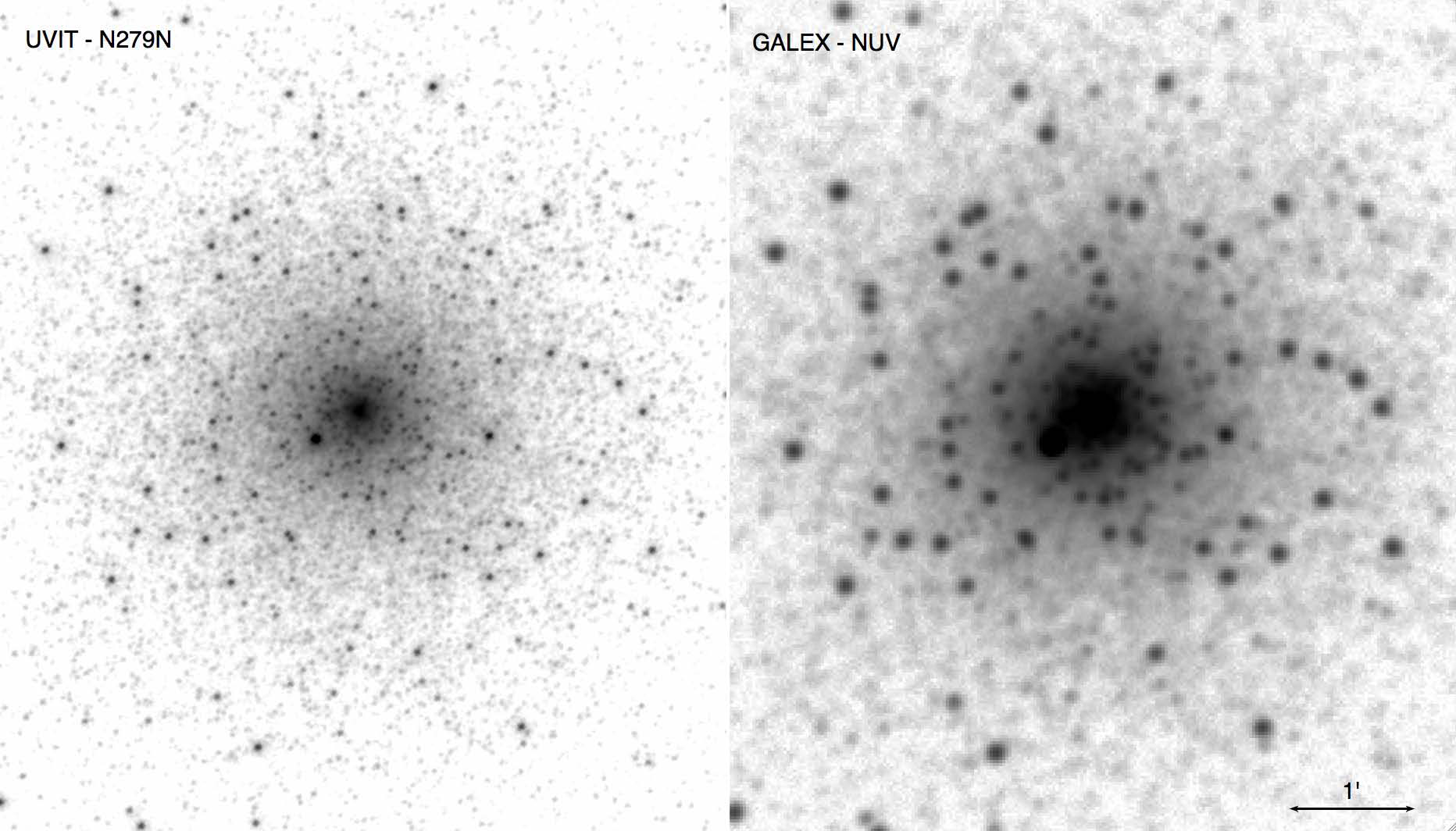}
\caption{Comparison of NUV images for NGC 1851 taken with UVIT and GALEX (left and right panels, respectively).
The UVIT image was constructed from a total integration time of 12234.35 sec in the F279N filter. North is up and east is to
the left. Only a subset of the full image is shown in each case.}
\end{center}
\end{figure*}

\begin{figure*}[h]
\begin{center}
\includegraphics[scale=0.2]{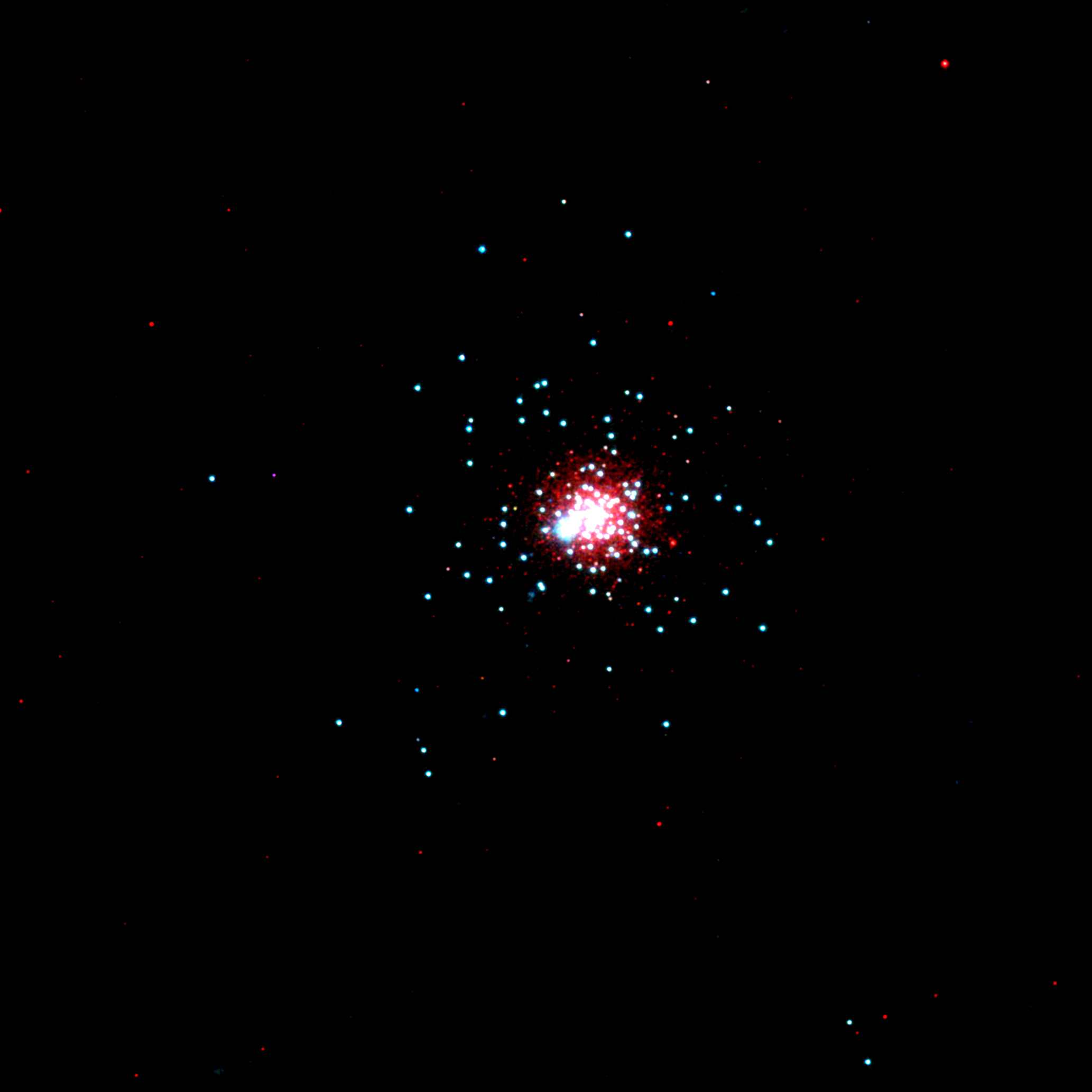}
\caption{This color image is made using the images obtained in two filters in the FUV and one filter in the NUV.
}
\end{center}
\end{figure*}

\begin{figure}[h]
\begin{center}
\includegraphics[scale=0.75]{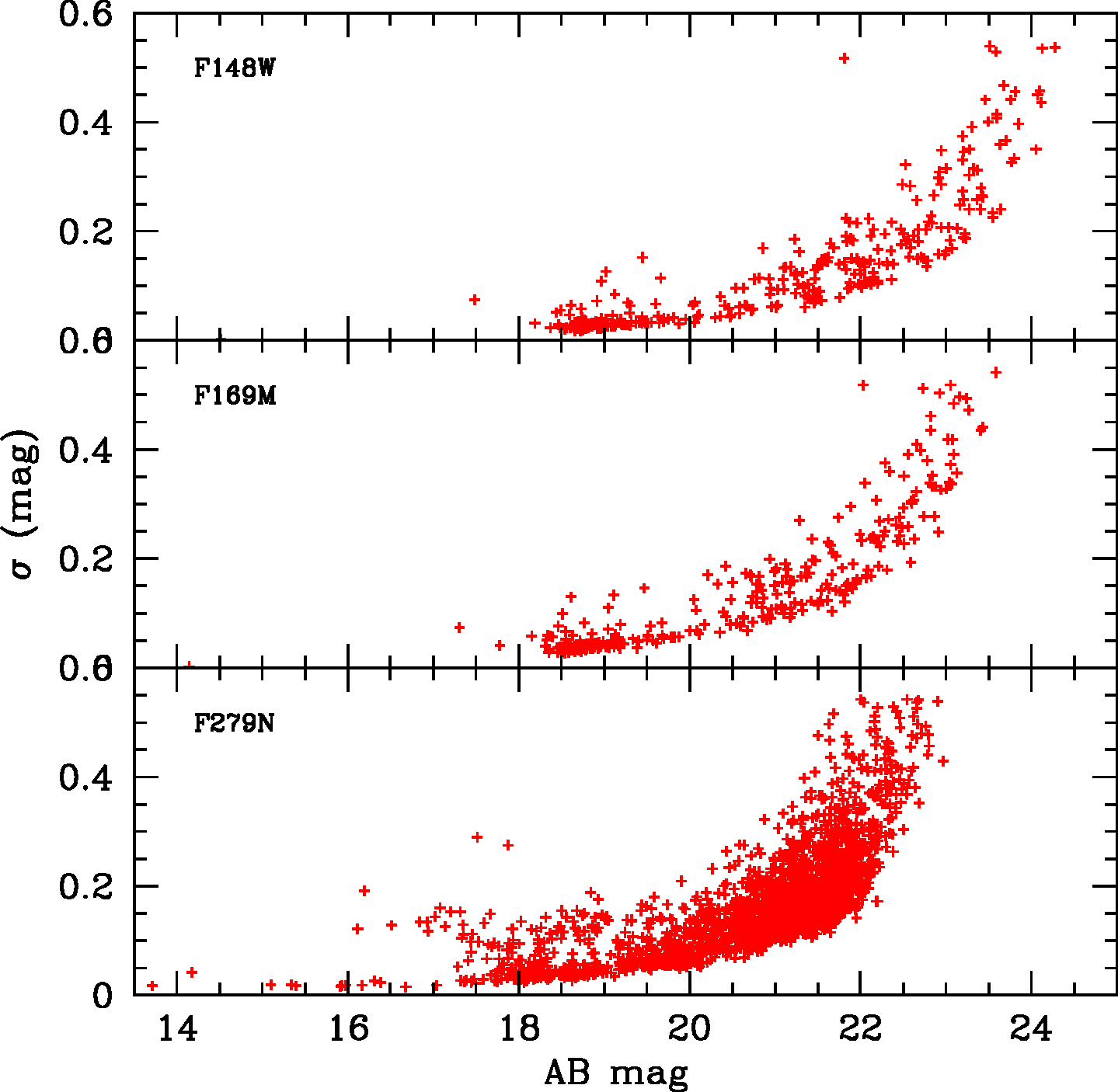}
\caption{Photometric errors as a function of magnitude for our UVIT observations of NGC 1851. From top to bottom, the panels 
show results for the F148W, F169M and N279N bandpasses, respectively.}
\end{center}
\end{figure}

Aperture photometry was performed 
on the images to estimate the counts after correcting for the background.
The point spread function (PSF) for isolated stars were constructed and was applied to all the detected stars, in order to
perform PSF photometry. A curve-of-growth  analysis was carried out to estimate aperture corrections in each filter, which
were then applied to the PSF magnitudes. These magnitudes were calculated after applying saturation corrections.
The above procedures were performed by two teams separately and consistency checks were
performed. The details of observations, photometry and calibrations are provided in Table 1.  The data presented in this study will be made available electronically.
Figure~4, we show our
photometric errors as a function of PSF magnitude.
Although we detect stars as faint as 23 magnitude in all filters, we have limited the comparison to the models
to 22 mag in FUV and 20.5 in NUV filters. This effectively puts an upper limit of 0.2 mag on our photometric errors. 
Of course, as Figure~4 shows, brighter stars will have photometric errors that are considerably smaller than this.

\begin{table}[h]
\centering
\caption{Details of photometry are listed. The first column lists out the filters used in this study, the second and the third columns list the mean wavelength and the passband of the filters, the fourth column lists the exposure time, followed by the zero-point magnitude, aperture correction (mag), number of detected stars and the Unit  conversion factor (to get flux in $ergs~cm^{-2}~s^{-1}~\AA^{-1}$) for the filters. }
\begin{tabular}{cccccccc}
\hline
Filter  & $\lambda_{mean}$  & $\Delta \lambda$  & $T_{\rm exp}$ & $m_{\rm zp}$ & Aperture Correction & $N_*$&UC \\
&(\AA) &(\AA) & (sec) & (mag) & (mag) &  &\\
\hline
F148W  &1481&500&~6982.13 & 18.00  & 0.15 & 330 &3.09E-15\\
F169M  &1608&290& ~5274.15 & 17.45 & 0.15 & 306 & 4.39E-15\\
N279N  & 2792&90& 12234.35 & 16.46 & 0.35 & 2298& 3.50E-15\\\hline
\end{tabular}
\end{table}

\section{The UV and Optical Color Magnitude Diagrams}
\label{sec:cmd}

The UVIT images, with their excellent resolution, allow us to resolve most stars in the central regions of the cluster (in
the case of FUV images). The NUV images, on the other hand,
do still exhibit crowding in the central regions. The FUV filters (F148W and F169M) are broad and medium 
band filters, whereas the
NUV filter (N279N) is a narrow band filter that is centered on the MgII line (2800\AA).  
 In order to show spectral coverage of the UVIT filters, we have plotted the filter effective areas on two typical BHB and RHB spectra in figure~5.
The parameters corresponding to the plotted spectra are mentioned in the figure caption.

To identify and classify the stars detected with UVIT into various evolutionary phases, we used the HST/ACS survey 
data  for Galactic globular clusters \citep{Sarajedini2007} 
to cross match our stars inside a region of diameter $\sim$ 4\farcm0. We have converted the V magnitude to the AB magnitude system, to make
it similar to the UVIT magnitude system. These stars were found to primarily belong to the HB of the cluster.
The cross matched stars are shown in the CMD (Figures 6 and 7).
The identified stars are separated into the horizontal part of the BHB (H-BHB), vertical part of the BHB (V-BHB), 
red part of the HB (RHB) and the post-AGB stars, according to their location in the optical HST-ACS CMD. 
On the AB magnitude scale, V-BHB stars are those with $(V-I) < -0.38$, which is the starting point of this nearly vertical sequence.
For comparison, BHB stars are those having $-0.38 \le (V-I) < -0.17$ while RHB stars have $0.02 < (V-I) < 0.22$.
The RR Lyrae stars detected in this study are also shown in these figures. 


\begin{figure}[h]
\begin{center}
\includegraphics[scale=0.3]{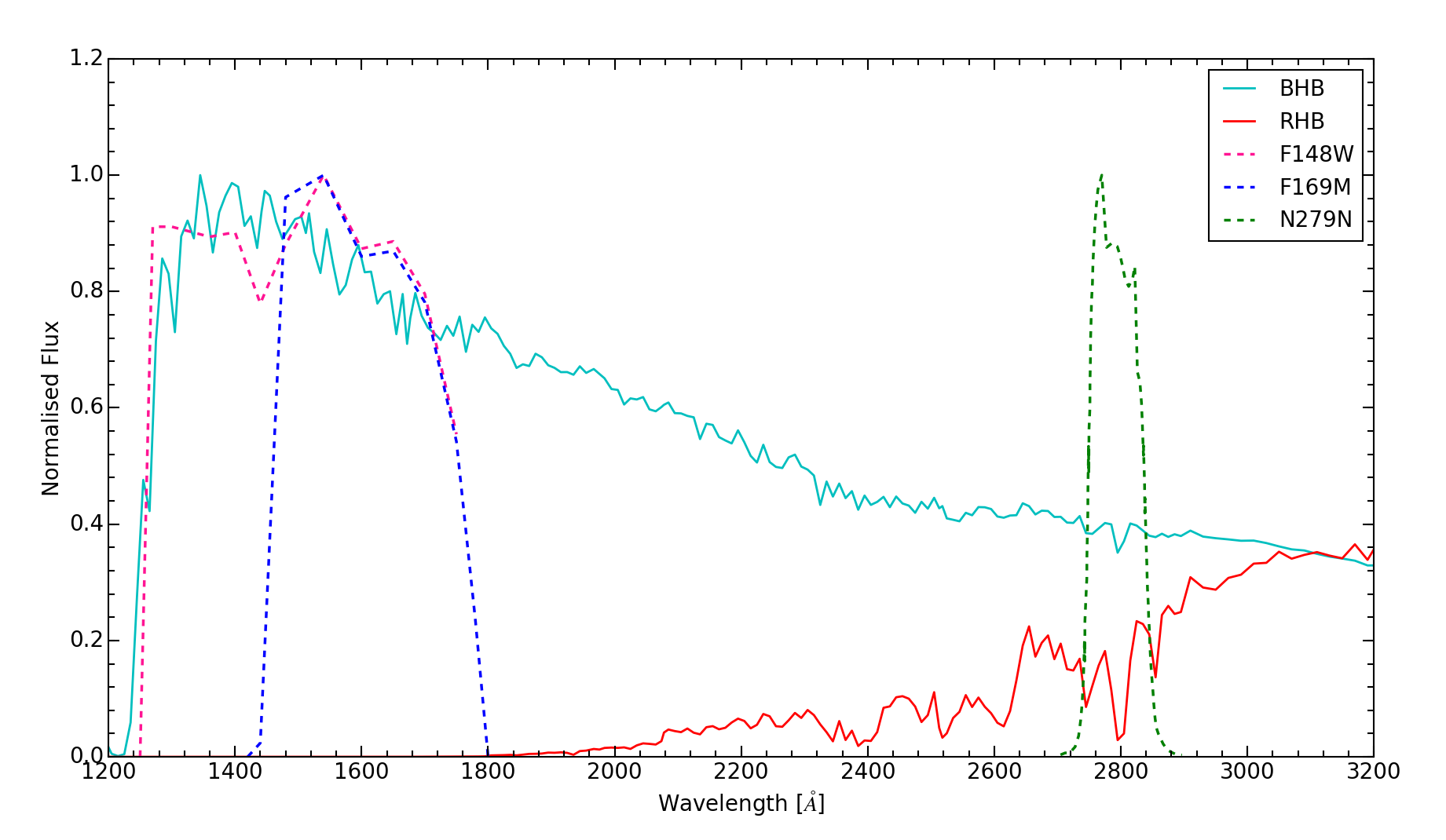}
\caption{We have shown effective areas of  the UVIT filters used in this study on a typical RHB and BHB spectra. The parameters of the BHB spectrum are
T$_{eff}$ = 10790K, L/L$_\odot$ = 1.776, Log(g) = 3.513, [Fe/H] = -1.2, Age = 12 Gyr and Y$_{ini}$ = ,0.28 and those for the RHB spectrum are
T$_{eff}$ = 5610K, L/L$_\odot$ = 2.713, Log(g) = 2.492, [Fe/H] = -1.3, Age = 12 Gyr and Y$_{ini}$ = ,0.23.}
\end{center}
\end{figure}

The isochrones are generated using the Flexible Stellar Population Synthesis (FSPS) models of \cite{Conroy2009}. These models include 
isochrone tables for the Bag of Stellar Tracks and Isochrones (BaSTI, \cite{Pietrinferni2004} and Padova isochrone sets \citep{Marigo2007, 
Marigo2008}. We used  both models and filter effective area curves for the UVIT filters (\cite{Tandon2017}, after incorporating the
in-orbit calibrations) to generate magnitudes 
in our UVIT filters. The isochrones shown in Figures 6 and 7 refer to the Padova models, for a metallicity of [Fe/H] = $-$1.2~dex 
\citep{Kunder2013}, a distance modulus of 15.52 mag and an assumed age of 10 Gyr \citep{Cassisi2008}. We point out that the HB stars no longer 
define a horizontal sequence in the UV CMD; note that even the H-BHB gets stretched in magnitude when compared to the V-BHB.

In the right panel of Figure~6, we show all stars detected in the FUV (188 stars), in the F148W vs (F148W$-$N279N) CMD. In the left panel 
we show the same stars in the 
optical CMD based on HST photometry. We are able to detect V-BHB and H-BHB stars, and five RR Lyrae
stars, in the central region of the cluster but are unable to detect RHB stars in the FUV. Hence, these stars are not 
present in the  observed CMD. The H-BHB stars are found to closely match
the isochrone, with the V-BHB stars located at the brighter end of the sequence. The V-BHB stars 
tend to show relatively more scatter, when compared to the H-BHB stars. We also note the presence of a few H-BHB stars, 
as bright as the V-BHB stars. In summary, we do not see significant differences between the V-BHB 
and H-BHB populations, except for a large scatter in the distribution of the V-BHB stars.
The RR Lyrae stars are suggestive of belonging to this population. We detected two Blue Stragglers which are also shown
in the figure. In Figure~7, we show the F169M vs. (F169M$-$N279N) CMD. The corresponding 
optical CMD based on HST photometry, is similar to that shown in the left panel of figure~6. 
The principle features of the CMD are similar to those in the Figure~6, suggesting that the
V-BHB and H-BHB populations appear to be similar. We also detected five blue stragglers that are shown as blue dots. In both the CMDs,
the B-BHB is stretched to a large magnitude range, while the V-BHB stars scatter over a smaller range in magnitude.  
\begin{figure}[h]
\begin{center}
\includegraphics[width=18cm,height=8cm]{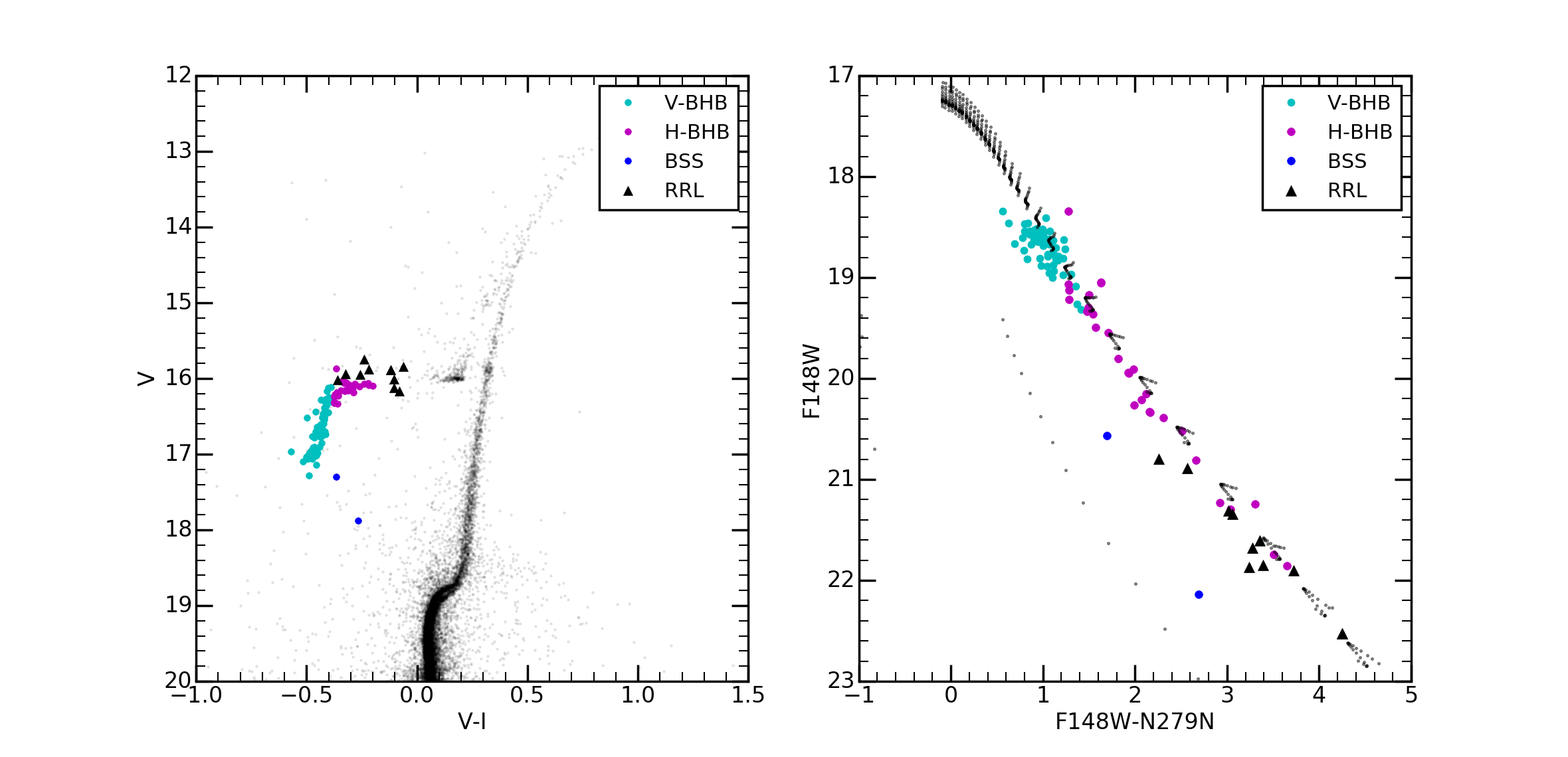}
\caption{CMD for NGC 1851 based on F148W and N279N photometry from UVIT (right panel). For comparison, we
overlay a Padova model isochrone with 10 Gyr and $[$Fe/H$]$ = $-$1.2~dex generated using the FSPS models in the UV CMD which are shown as black dots.
The panel on the left shows the optical CMD based on HST/ACS photometry.
Stars detected by UVIT have been separated into five components: vertical-blue horizontal branch (V-BHB), 
horizontal-blue horizontal branch (H-BHB), red horizontal branch (RHB) (not shown in this CMD), blue stragglers stars (BSS), and RR Lyrae stars (RRL).}
\end{center}
\end{figure}
\begin{figure}[h]
\begin{center}
\includegraphics[scale=0.34]{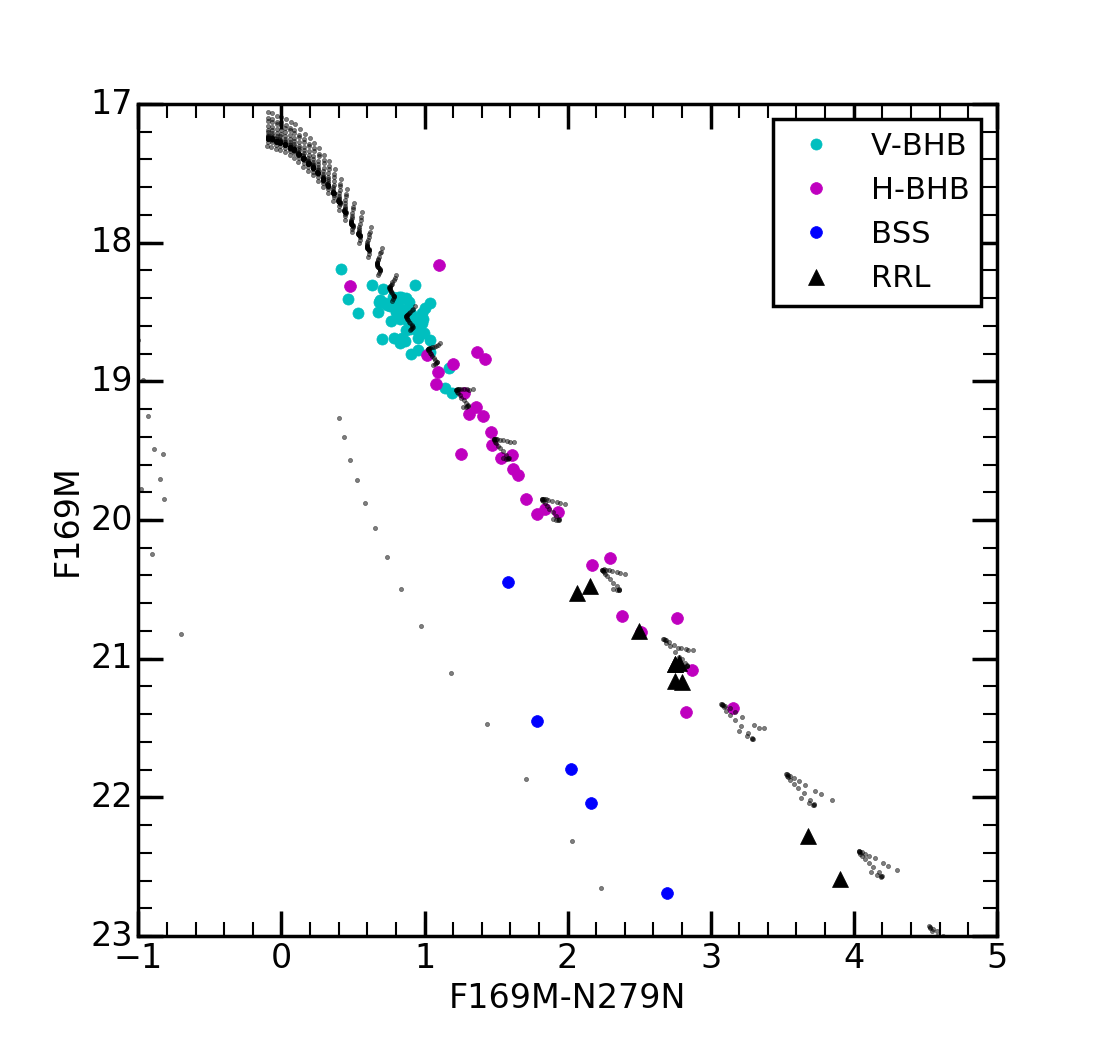}
\caption{CMD for NGC 1851 based on F169M and N279N photometry from UVIT. For comparison, we
overlay a Padova model isochrone with 10 Gyr and $[$Fe/H$]$ = $-$1.2~dex generated using the FSPS models in the UV CMD which are shown as black dots. The various symbols show stars belonging to different evolutionary sequences in the CMD, as described in Figure~6.
}
\end{center}
\end{figure}

\begin{figure}[h]
\begin{center}
\includegraphics[scale=0.25]{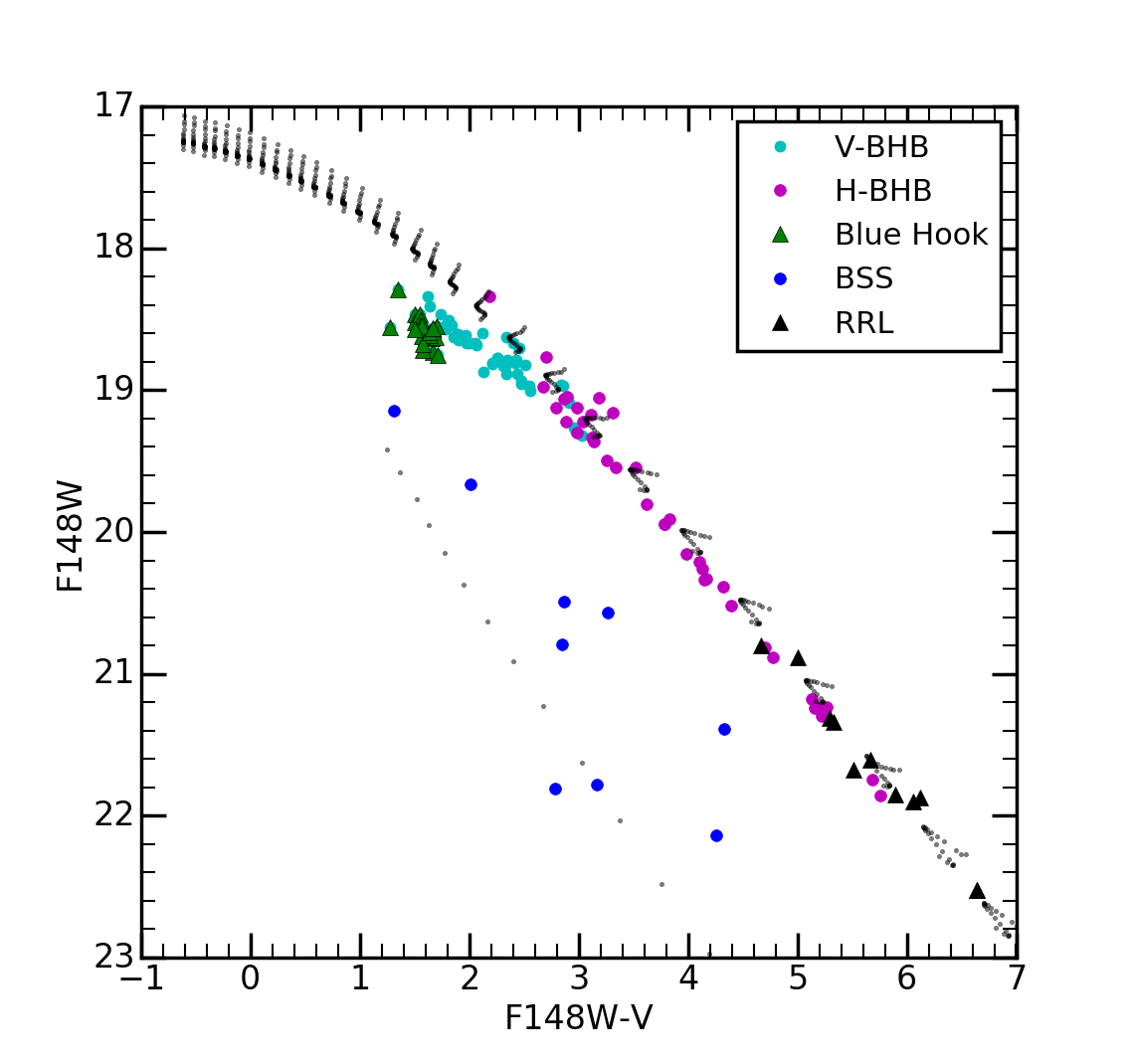}
\includegraphics[scale=0.25]{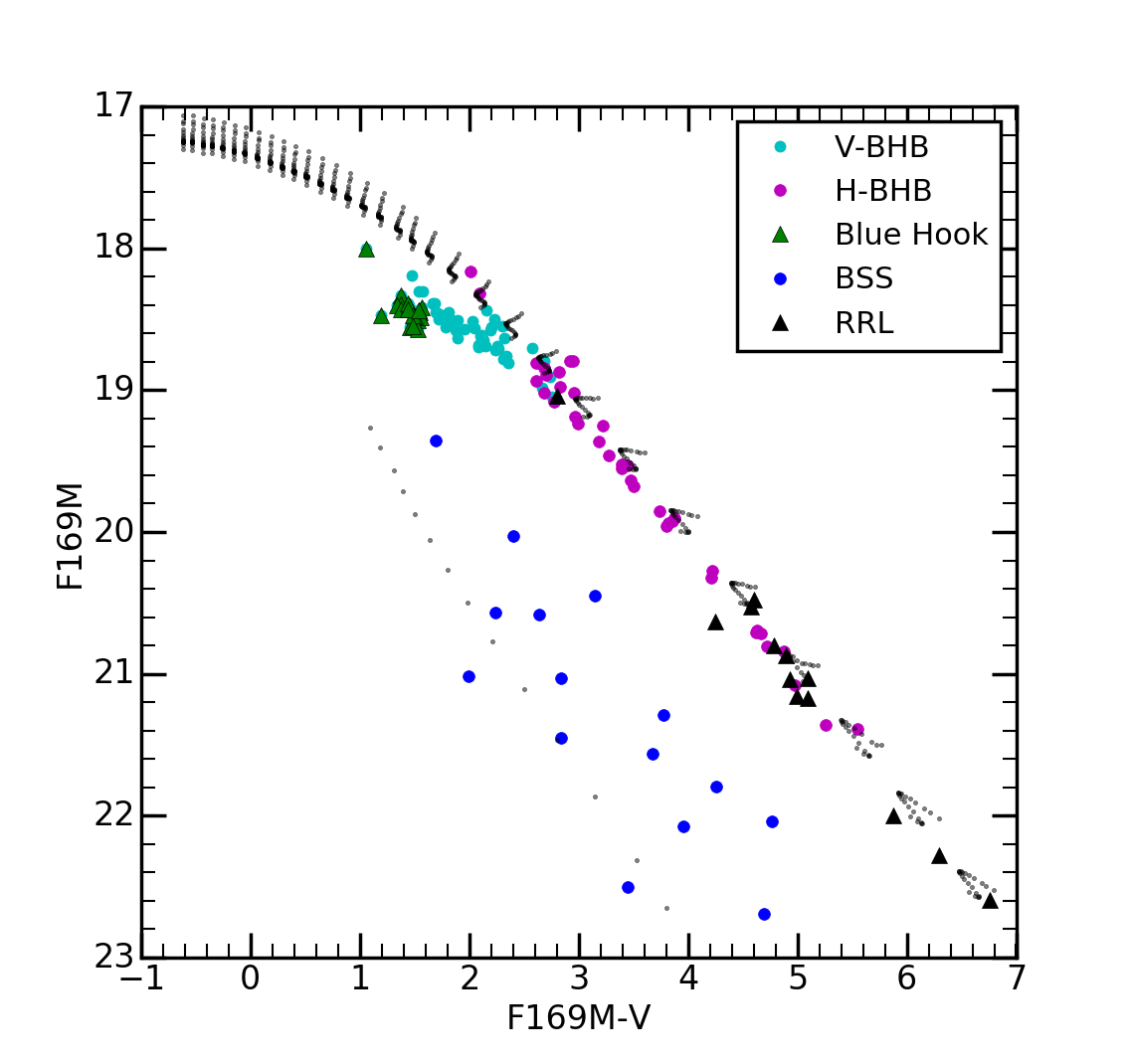}\\
\includegraphics[width=16.3cm, height=6.7cm]{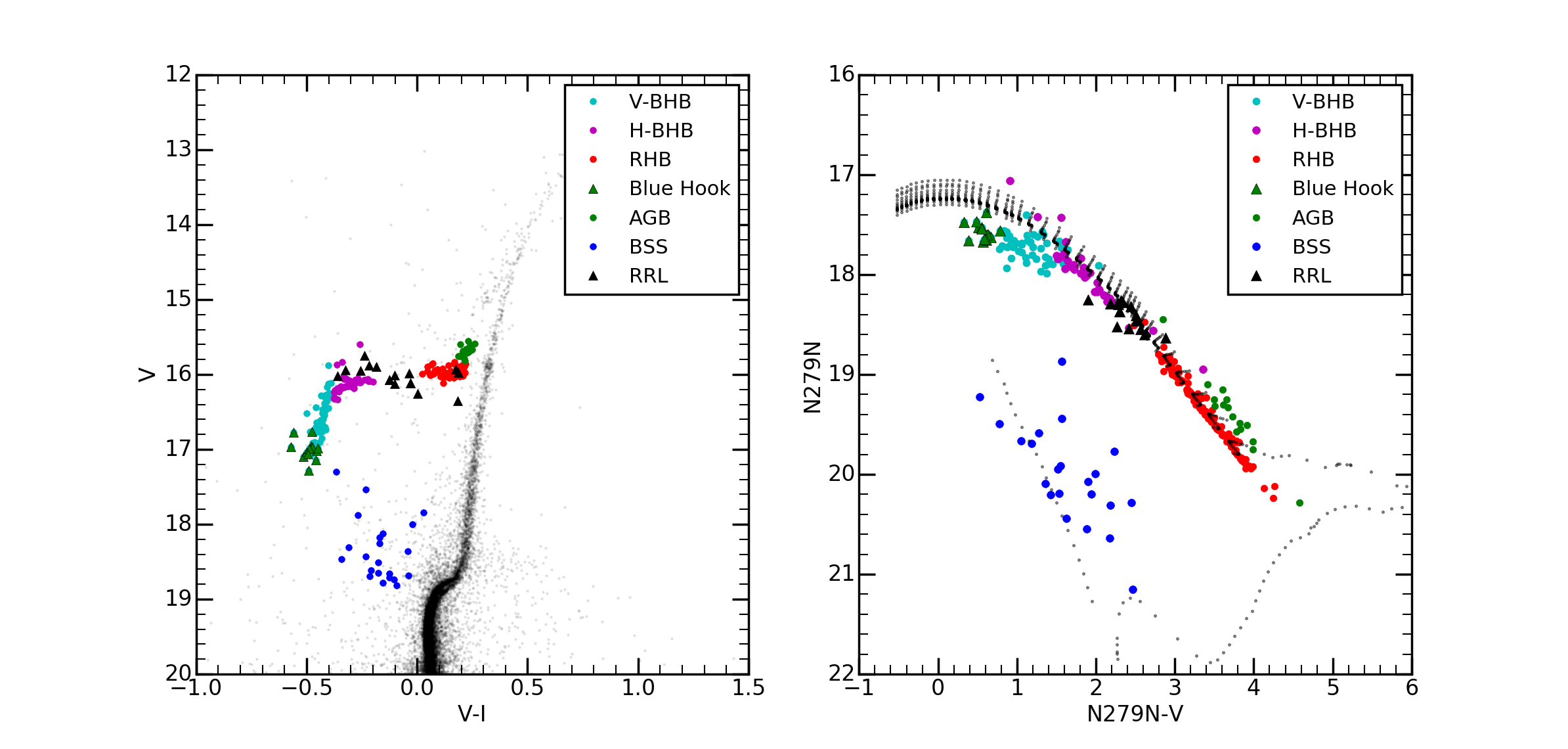}\\
\caption{UV CMDs for NGC 1851 after cross-matching HST/ACS data to UVIT data in the F148W, F169M and N279N filters. FUV CMDs are shown in the upper panels. NUV and the corresponding optical CMDs
are shown in the lower panels. For comparison, we
over plot a Padova model isochrone with 10 Gyr and $[$Fe/H$]$ = $-$1.2~dex generated using the FSPS models in the UV CMDs which are shown as black dots. The various symbols show stars belonging to different evolutionary sequences in the CMD, as 
described in Figure~6.}
\end{center}
\end{figure}

In Figure~8, we combine the UVIT and HST photometry to plot UV-optical CMDs. We have shown the
F148W vs (F148W$-$V) (top left panel), F169M vs (F169M$-$V) (top right panel) and N279N vs (N279N$ - $V) (bottom right panel), 
along with the cross matched stars in the optical CMD using the HST data (bottom left panel). In the UV-optical CMDs in the top
 panels, we detect the V-BHB, H-BHB and some RR Lyrae stars. In the case of bottom panel, we detect the full 
HB along with the RR Lyrae stars. The HST CMD shows all the detected stars in the NUV, whereas only a subset is detected in the FUV.
The isochrones with the same age and metallicity (as in Figures 6 and 7) 
are shown here, for the corresponding filters. We also detect some blue straggler stars. It can be seen that, 
in all the three UV-optical CMDs, the V-BHB stars deviate from the H-BHB stars. A few of the
H-BHB stars are found to continue the H-BHB slope and end up brighter than the V-BHB stars.
The blue hook feature can be clearly identified at the bright end of the HB in the FUV CMDs (upper panels), 
suggesting the presence of the blue hook stars at the end of the V-BHB sequence. 
The H-BHB stars are found to be in a sequence, which could be an extension of the RHB stars, but we do detect 
some variation (bottom panel), suggesting that the H-BHB branches off with a shallower slope.
As expected, the RR Lyrae stars are located between the H-BHB and RHB stars. 
A careful inspection reveals that the V-BHB stars can be seen
to branch off from the sequence of H-BHB stars, with a bit of mixing of stars near the point of deviation. 
We also see that some RHB stars and H-BHB stars are mixed together, even though these groups are well separated in 
the optical CMD. The RHB stars are found to have a very tight sequence, with very little scatter, when compared to the 
V-BHB and B-BHB stars. We find the AGB stars to be brighter than the RHB and separated from it (bottom panel). 

In all panels, the V-BHB stars are found to be fainter than the isochrone, whereas the B-BHB stars are almost 
aligned with the isochrone. This may point to some differences between these two types of stars. 
The RHB stars are found to extend below the magnitude limit of the HB, as delineated by the isochrone (bottom panel). 
This could suggest that either the RHB is metal-rich or younger than the over plotted isochrone. 
We find that there could be differences among the V-BHB, H-BHB and the RHB stars, with the V-BHB and
RHB stars possibly belonging to two different populations.  The differences could be due to a difference in age, metallicity, CNO abundance or mass loss, as mentioned
as possible mechanisms by previous studies on this cluster. 

Finally, we cross-matched the UVIT detected stars with ground-based V,I photometry provided by Peter Stetson (private
communication). In Figure~9, we present the CMD obtained by cross matching with the ground data, where the UVIT CMD is on 
the lower right panel and the corresponding optical CMD is in the left panel. It can be seen that the number of cross matched stars from the ground data (59 stars) is much less than stars from the HST data (254 stars), which covers the inner $0\farcm5-4\farcm0$ region. 
Nevertheless, we detect all the three types of HB stars in the ground photometry (28 RHB, 6 HBHB, 13 VBHB ), as
well as 12 blue straggler stars. By comparing the inner data from the HST and the outer data from the ground, it can be seen that most of the HB stars are located within a region of diameter $0\farcm5-4\farcm0$. Indeed, 
only a modest fraction of the detected stars (47/232 $\sim$ 20\%) are found outside this region. 

\begin{figure}[h]
\begin{center}
\includegraphics[scale=0.3]{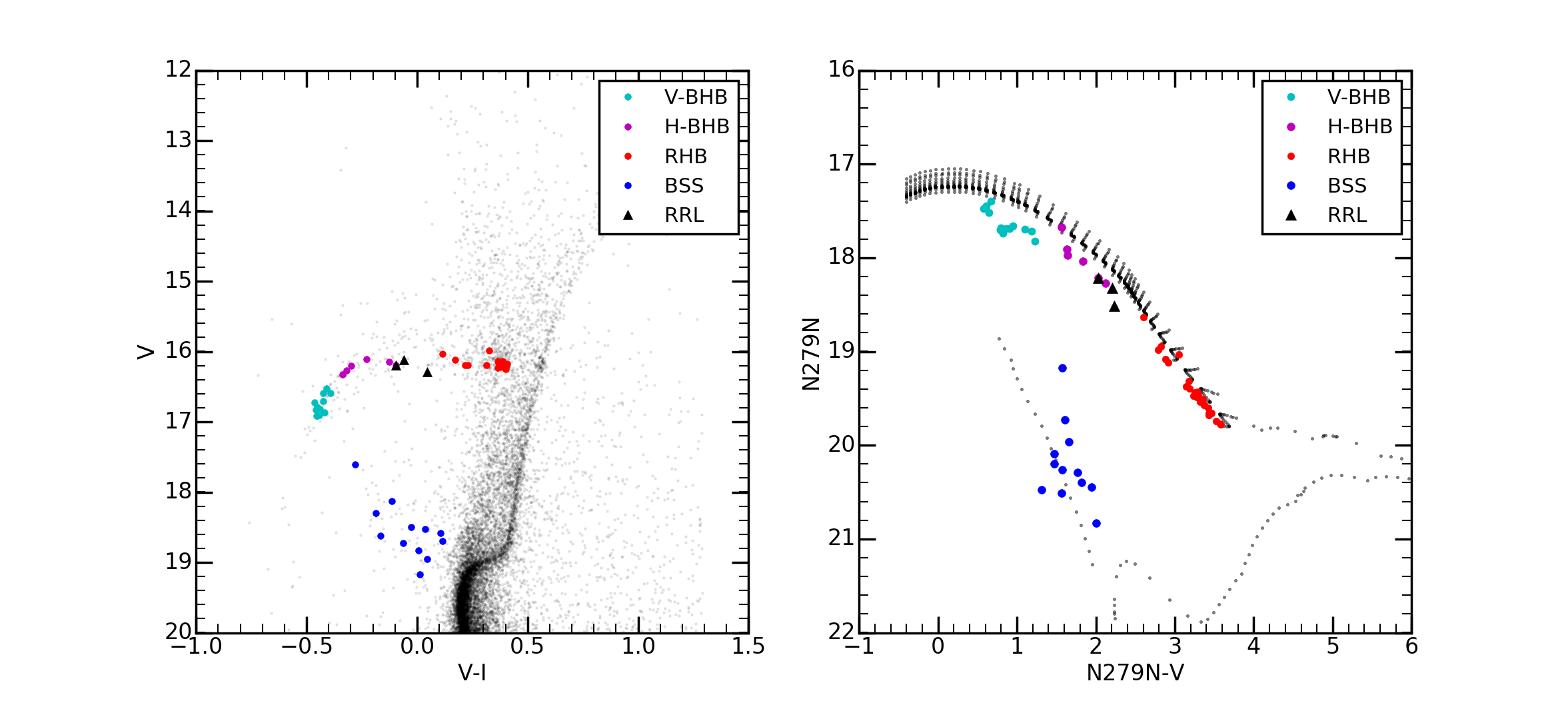}
\caption{UV-optical CMDs obtained by cross-matching ground-based data with UVIT data in the N279N filter. For comparison, we
overplot a Padova model isochrone with 10 Gyr and $[$Fe/H$]$ = $-$1.2~dex generated using the FSPS models in the UV CMD which are shown as black dots. The various symbols show stars belonging to different evolutionary sequences in the CMD, as described in Figure~6}
\end{center}
\end{figure}

\section{Characterizing the HB Population}
\label{sec:analysis}

We have attempted to characterize the HB stars  by comparing the UV-optical CMDs with predictions from the stellar evolutionary models. 
Before proceeding, we note that \citep{Gratton2012} studied the HB population in NGC 1851
using medium resolution spectroscopy and found that, on average, the RHB stars have [Fe/H] $= -$1.14$\pm$0.01~dex from Fe I lines
and [Fe/H] = $-$1.20$\pm$0.01~dex from Fe II lines. They also measured a He abundance of Y=0.291$\pm$0.055, and estimated
an initial He abundance of about Y = 0.248 with mild He enhancement thereafter. A large initial He abundance was deemed unlikely.

Figure~8 suggests that the isochrone fits the HB in general well, except for the deviation noticed for the V-BHB stars. The isochrone also suggests fainter extension to the RHB.
These deviations could suggest that one or more parameters corresponding to the isochrone plotted, might vary among the detected HB stars.
We also used the BaSTI models to generate isochrones with different ages and abundances, instead of the Padova models. 
As we need to fold in the UVIT filters into the FSPS models, we are able to change
only the age and metallicity of the isochrones, keeping the other parameters, such as Y$_{ini}$ constant. 
In Figure~10, we show isochrones for two ages, where in each plot, we have kept the age constant
and varied the metallicity as shown in the figure. These isochrones have  Y$_{ini}$ = 0.25.
The figure shows how the isochrones for the HB stars change in the UV with age and metallicity, as per the predictions from BaSTI models. The isochrones shown
in both the figures could be used to derive the range of age and metallicity.  With the decrease in the value of [Fe/H], the extent of
the HB shrinks such that the red end of the HB gets bluer. The presence of an extremely
extended HB, as seen in this cluster, can be fitted using isochrones with [Fe/H] $\sim$ $-$1.0. The fitted value of [Fe/H] is, to within the errors, in good agreement with 
the values obtained by \cite{Gratton2012}.

\begin{figure}[h]
\begin{center}
\includegraphics[scale=0.35]{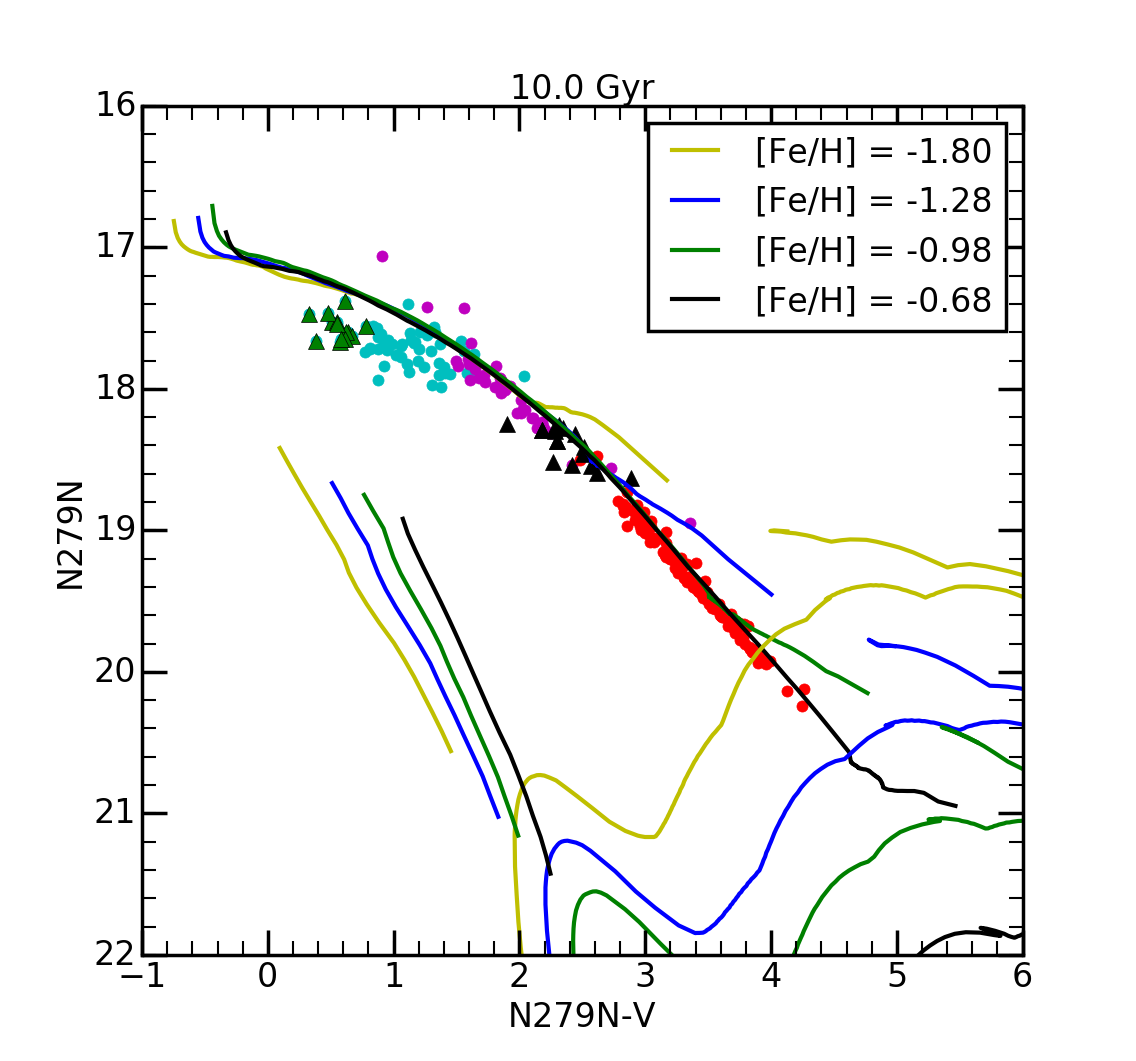}
\includegraphics[scale=0.35]{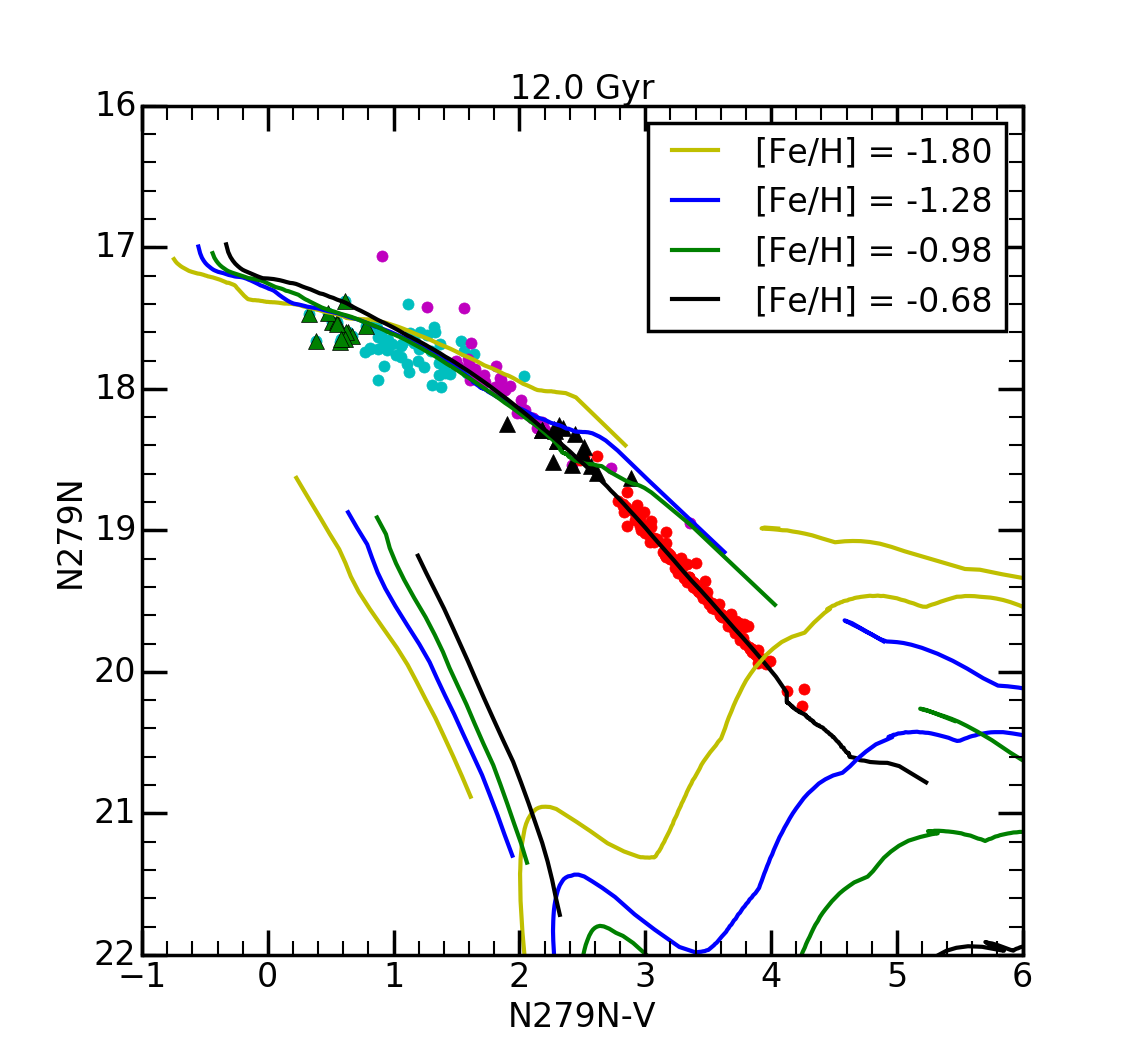}
\caption{NUV-optical CMD for NGC 1851 based on UVIT imaging in the N279N filter. BaSTI Isochrones are shown for four different values metallicities, as indicated
in the legend, and two assumed ages: 10 and 12.0 Gyr (left and right panels, respectively).
}
\end{center}
\end{figure}

From the upper panel of figure~10, it is clear that the RHB can be fitted using an isochrone of age 10 Gyr, for the above value of [Fe/H], 
whereas none of the isochrones in the bottom figure fits the RHB. The bottom figure shows that the fit of the RHB cannot be reproduced by
an increase in age. None of the isochrones in the top and bottom panel are able to fit either the V-BHB or the H-BHB. A close examination
of the bottom figure suggests that the isochrone shown in green points, corresponding to 12 Gyr, is the closest fit observed for the H-BHB 
stars.  Thus, we find that an  age range of up to 2 Gyr might be required to fit  the full HB stars,  for a constant Y$_{ini}$ and metallicity. 
In figure~11, we have shown the CMD with isochrones of two ages (10, 12 Gyr) for Fe/H] $\sim$ $-$1.0 and Y$_{ini}$ = 0.25, for the UV-optical CMDs.
These are the parameters we are able to
derive for the HB stars using the BASTI isochrones. In summary, we detect the presence of two age groups in 
the HB of NGC 1851, using the BaSTI isochrones. We find that the RHB stars in the cluster could belong to a relatively younger population (10 Gyr), whereas the BHB 
stars could be up to 2 Gyr older.  We believe that there cannot be large difference in [Fe/H] among 
the HB stars. Recall that \cite{Carretta2011} and \cite{Gratton2012} suggested that the cluster has two populations, with an age difference of up to 2 Gyr. 

As noted above, the N279N filter is centered around the Mg~II line and therefore, the flux in this filter depends on the magnesium 
abundance of the HB stars. \cite{Gratton2012} found a higher abundance of magnesium for their BHB stars, in comparison to the 
RHB stars. The isochrones in this filter are expected to capture the details of variation in magnesium abundance 
across the HB. The fact that the FUV CMDs also suggest the deviation of V-BHB stars from the 10 Gyr isochrone suggests that 
the deviation may not be related to any residual variation in the Mg~II strength. This is confirmed from the bottom panel of figure~11, we have overlaid 
the isochrones on the FUV CMD. The  
CMD confirms that the V-BHB stars are reasonably well fitted by the 12 Gyr isochrone. The blue hook stars are fainter than 
the rest of the V-BHB stars and the isochrones do not reproduce this feature. Thus, we note that the blue hook stars appear to 
be fainter in the FUV, even though they have normal NUV flux. This may provide some specific clues to their structure and 
evolutionary phase. \cite{Dalessandro2011} presented the FUV-optical CMD of NGC 2808 and used it to confirm the 
multi-model distribution of stars along the HB. They found that the canonical models are not able to match the hot end of the HB.


\subsection{He Enrichment}
\label{sec:Herich}

The issue of He enrichment in this cluster has been discussed and contested for some time. 
The spectroscopic study of \cite{Gratton2012}, SGB study of \cite{Milone2008} as well as the HB simulations of \cite{Salaris2008}, did not find any significant 
enhancement in He.  \cite{Han2009}, \cite{Kunder2013}, suggested that He is enhanced in the BHB 
stars, apart from them being older. \cite{JooLee2013} also argued that He enhancement can explain the age effect, 
and therefore an apparent age difference may actually be due to differences in He content. 

In order to understand  the effect of Y$_{ini}$ on the HB morphology, we considered the He enhanced models. 
The synthetic HB models presented here are based on $Y^2$ stellar evolutionary tracks with enhanced initial He abundance \citep{Lee2015}. 
We adopt \cite{Reimers1977} mass-loss coefficient $\eta=0.5$ \footnote[1]{ In order to reproduce the HB morphology of inner halo MWGCs, we assume the age of the inner halo as 13.3 Gyr and the initial helium as Y$_{ini}$=0.23. Since the increased eta value has the same effect as the increased age or the increased initial helium abundance by decreasing the mean mass of HB stars, if we increase the initial helium or include helium spread in the inner halo, the eta becomes less than 0.5.} and the mass dispersion on HB stars $\sigma_{M}=0.015M_{\odot}$. We assume a fixed $\alpha$-element enhancement of $ [\alpha/Fe]=0.3$. We refer the reader to \cite{Chung2017} for more detailed descriptions of our synthetic HB models. Note that we did not consider the effect of the CNO abundance anomaly in the models. Therefore the variation of HB morphology is solely coming from the combinations of the metallicity, age, and the initial He contents.

In Figure~12, we present the synthetic HB plotted over the observed HB. 
The left panels show the fit to the observed HB in optical colors of ${\rm V-I}$.
We applied different values of Y$_{ini}$ for the bimodal HB morphology and splits on the sub-giant branch stars of NGC~1851 at the same time.
We assume helium enhanced population consists of slightly metal-rich population and fix the age of two population as 12~Gyr.
There are still issues on CNO abundance variation between two populations of NGC~1851, but as reported in \citep{JooLee2013} the 
variation is less than 0.1 dex which have almost negligible effects on HB morphology and splits of SGB stars.
Therefore, both of the HB morphology and the splits on SGB are caused by the difference in initial helium abundances.
In the right panels, we present the UV CMDs based on the same stellar parameters of the left panels.
The synthetic HB reproduces the RHB and the BHB stars very well,  though we see that the model predicts fainter BHB stars in V band and brighter BHB stars in N279N filter. The difference  between the model and the observed magnitude in the N279N is more than 3$\sigma$.
Thus, the synthetic CMD models with He enhancement suggest the following parameter combination for the HB population: an age of 12~Gyr, and a range in Y$_{ini}$ of 0.23 - 0.28 with a metallicity range of [Fe/H] $\sim$ $-$1.2 to $-$1.3. In particular, the parameter range for the RHB stars are $Y_{ini}$=0.23, 12~Gyr, and [Fe/H]=$-1.3$, whereas those for the BHB are, $Y_{ini}$=0.28, 12~Gyr, and [Fe/H]=$-1.2$. These parameters are estimated from the UV CMDs for the first time and are in good agreement with the previous estimated values of $Y_{ini}$, [Fe/H] and age, in the literature.


In order to explore population difference among the BHB stars, we used the two FUV filters in the UVIT to produce a 
F148W vs (F148W$-$F169M) CMD. Here the color term corresponds to the ratio of flux within the FUV band. The F148W 
is a FUV window with 1300-1800~\AA~coverage, whereas the F169M filter has 1450 - 1750~\AA~coverage. The CMD is 
shown in Figure~13. We have overlaid the BaSTI isochrones in the upper panel and the synthetic HB from $Y^2$ models 
in the lower panel. The photometric errors are also shown for better comparison. The H-BHB stars span 
a range of 3 magnitudes, whereas the V-BHB stars span a range of one magnitude, with the blue-hook stars co-located
with the brightest V-BHB stars. The color difference for the H-BHB stars is $\sim$ 0.3 mag, whereas for the V-BHB stars,
it is about $\sim$ 0.1 mag. The brightest of the H-BHB stars  have similar colors as the V-BHB stars. The HB stars are tend to be
bluer than the predictions, suggesting that they are relatively hotter. 
On the other hand, we do not see a significant color difference between the H-BHB and the V-BHB stars, more than that is 
suggested by the isochrone. The temperature of the brightest BHB stars, as per BaSTI as well as 
Padova models is found to be $\sim$ 12,000K. The $Y^2$ models reproduce the BHB stars better and the
best fit parameters for the H-BHB stars are 12 Gyr, Y$_{ini}$ = 0.28 and [Fe/H] $\sim$ $-$1.2 dex.  They also predict the
BHB stars to be more then 1 mag brighter than the observed BHB stars in the F148W filter. This needs further attention.
The estimated parameters are in overall agreement with \cite{Gratton2012}.  

\begin{figure}[h]
\begin{center}
\includegraphics[scale=0.35]{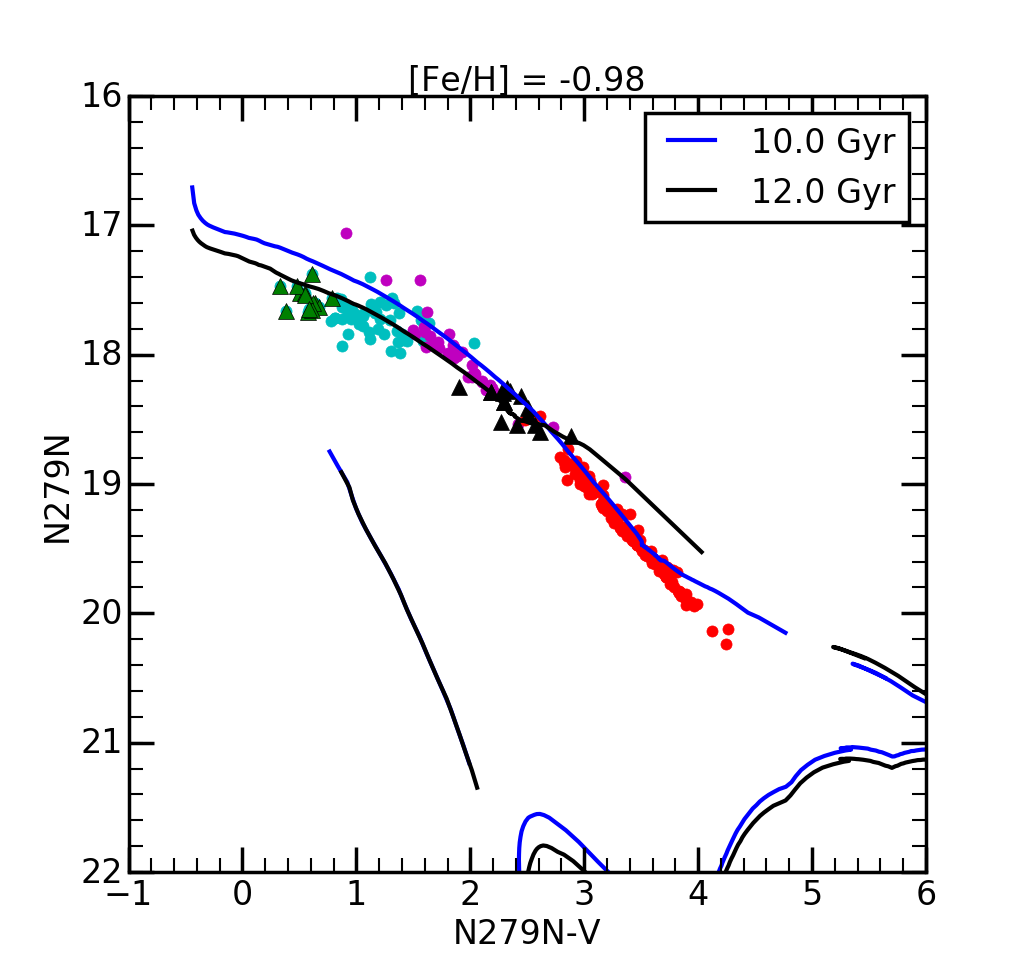}
\includegraphics[scale=0.35]{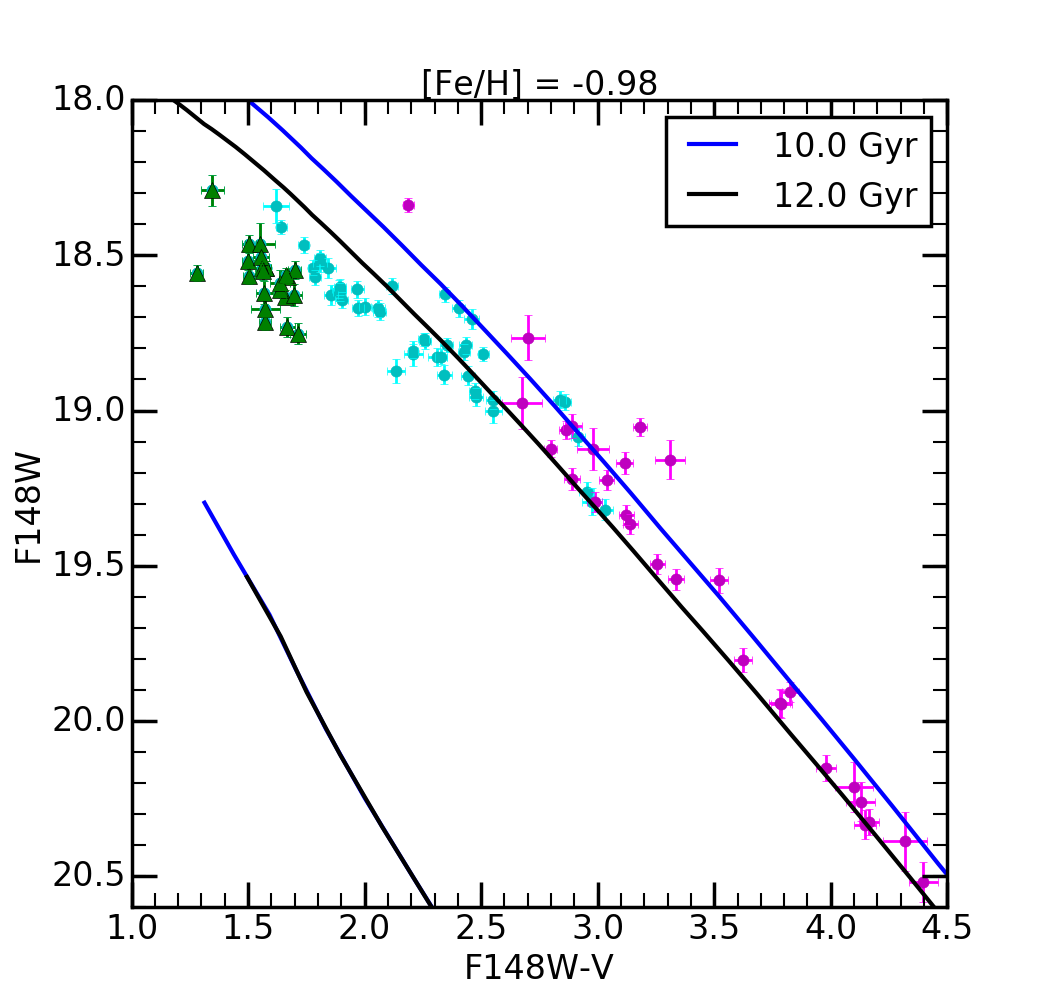}
\caption{NUV-optical CMD for NGC 1851 based on UVIT imaging in the N279N filter in the upper panel  and  FUV-optical CMD for NGC 1851 based on UVIT imaging in the F148W filter in the bottom panel. BaSTI Isochrones are shown for two different ages, as indicated in the legend, and a metallicity of [Fe/H] = $-$0.98~dex.}
\end{center}
\end{figure}

\begin{figure}[h]
\begin{center}
\includegraphics[scale=0.3]{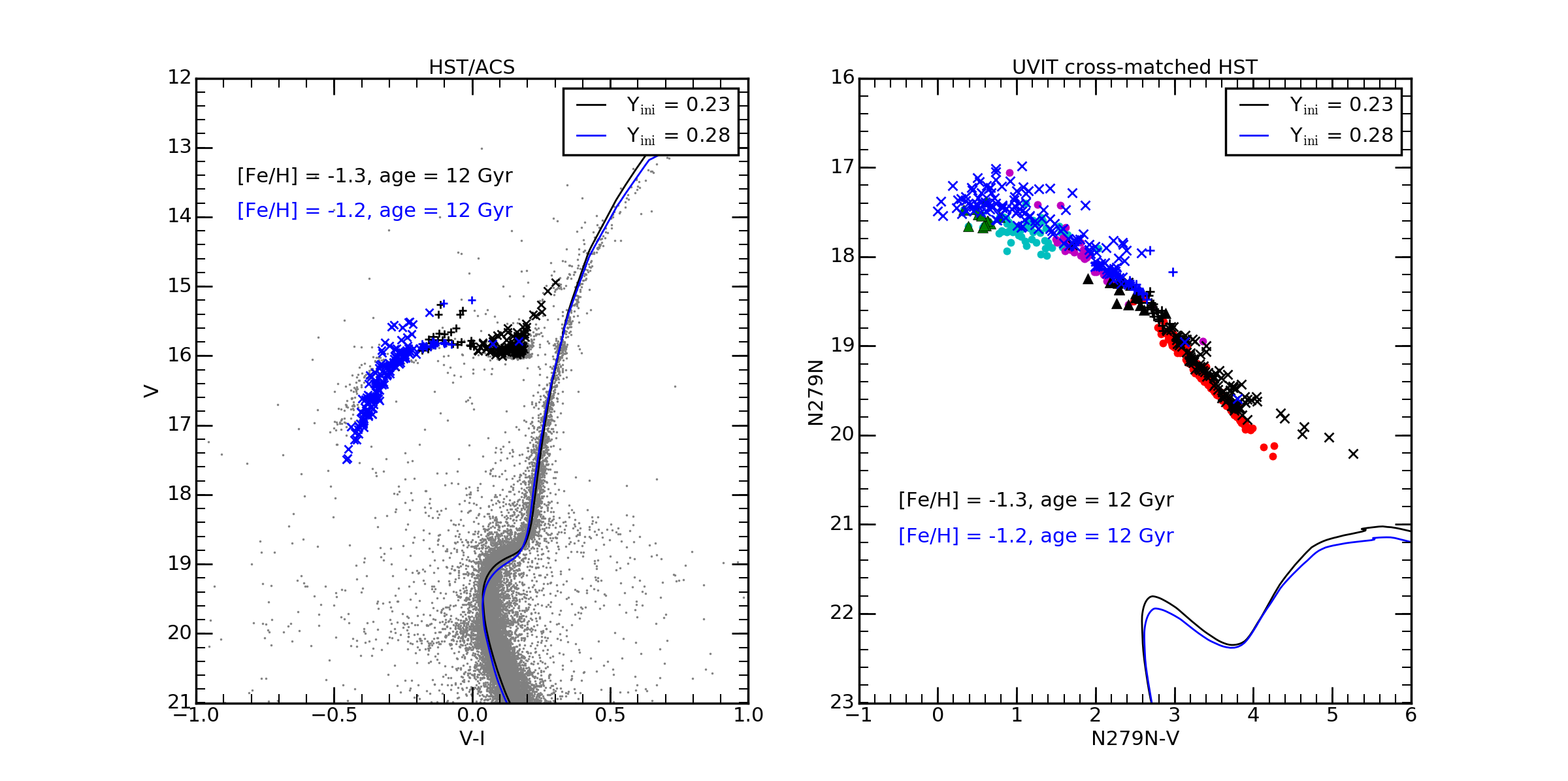}
\includegraphics[scale=0.3]{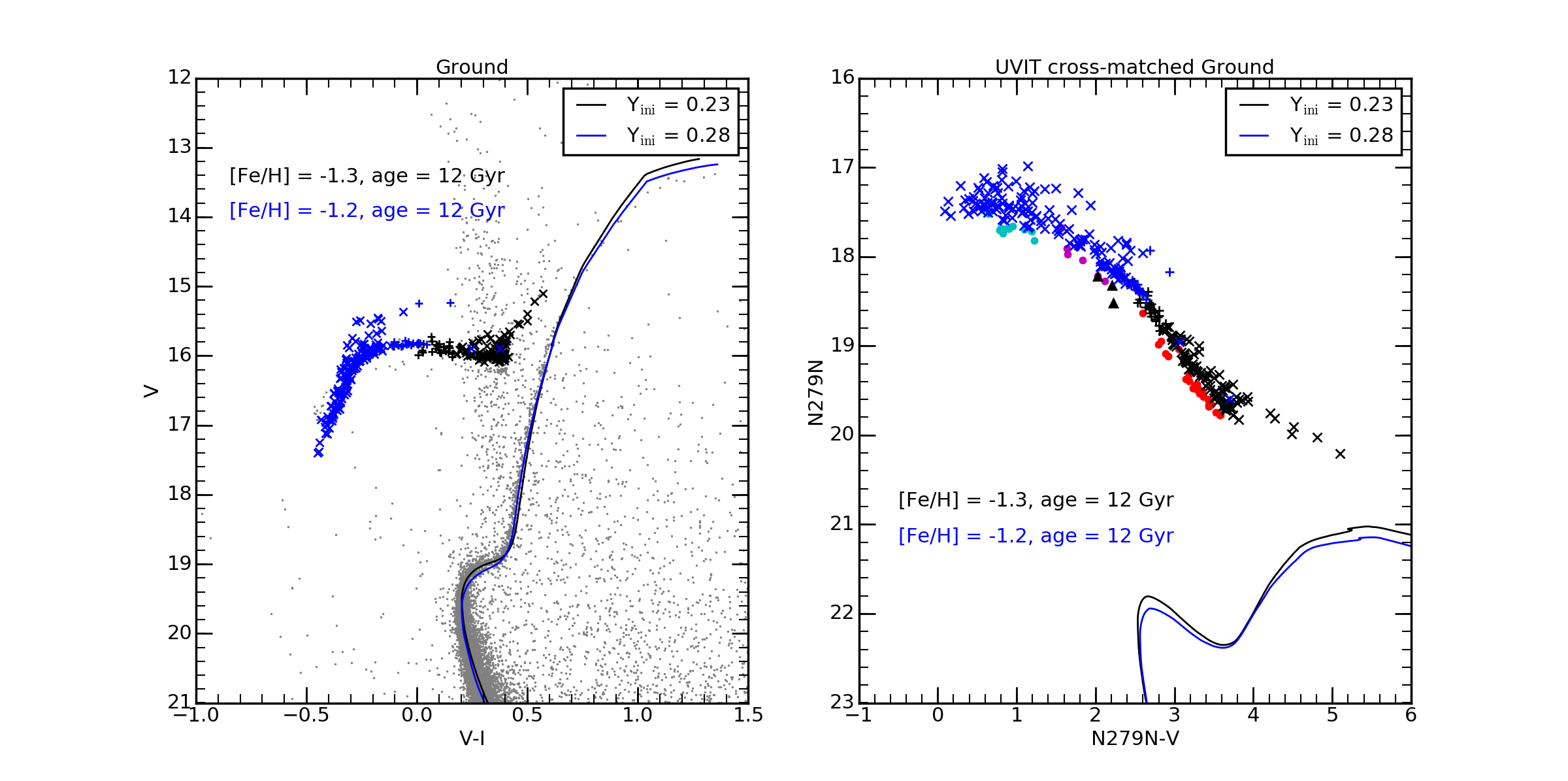}
\caption{Synthetic HB models from $Y^2$ stellar evolutionary tracks with enhanced initial He abundance,  Y$_{ini}$ 
\citep{Lee2015} are overlaid on the observed data
points. The upper panel shows the HST optical data in the left panel, and cross matched UVIT data in the right panel. The lower panel shows the ground based optical data
in the left panel and cross-matched UVIT data in the right panel. The crosses denote the generated synthetic HB, the lines indicate the $Y^2$  for the Ms turn-off and sub giant branch.}
\end{center}
\end{figure}

\begin{figure}[h]
\begin{center}
\includegraphics[scale=0.3]{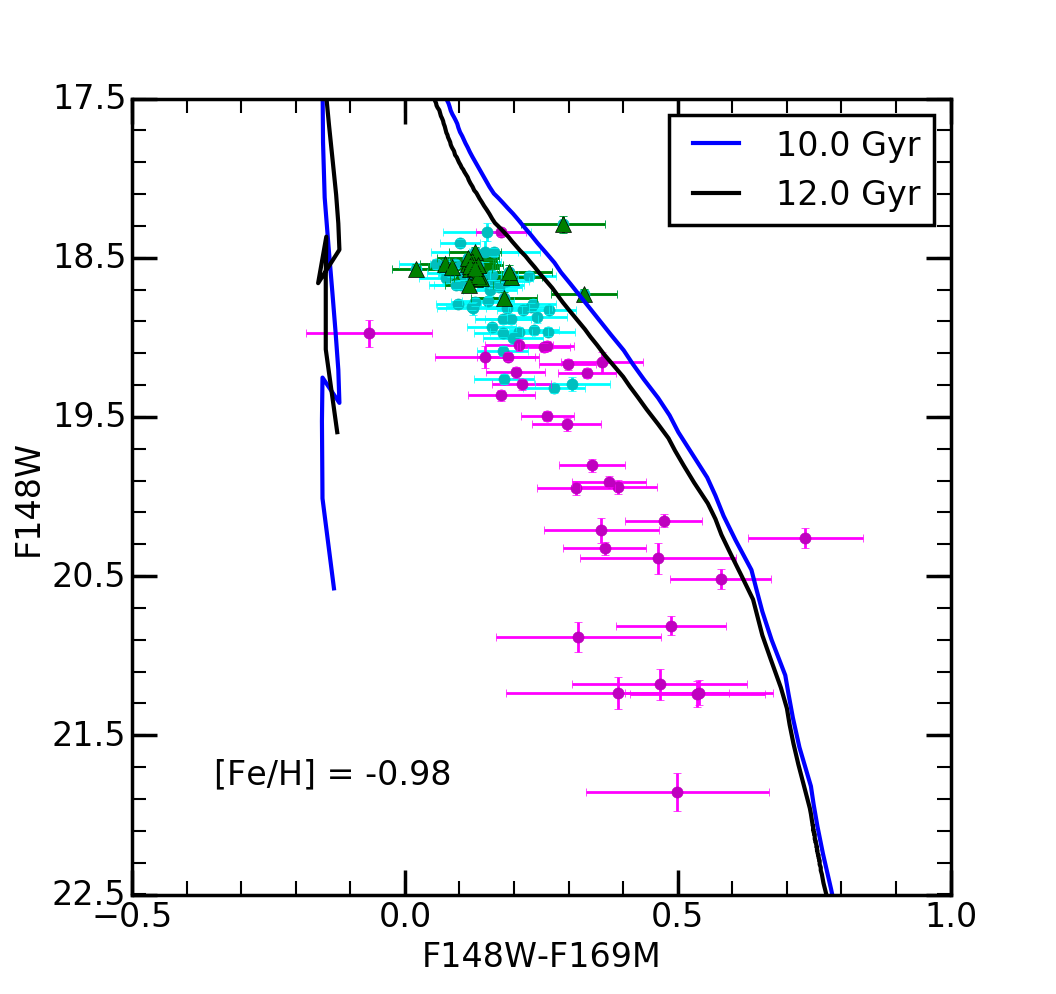}
\includegraphics[scale=0.32]{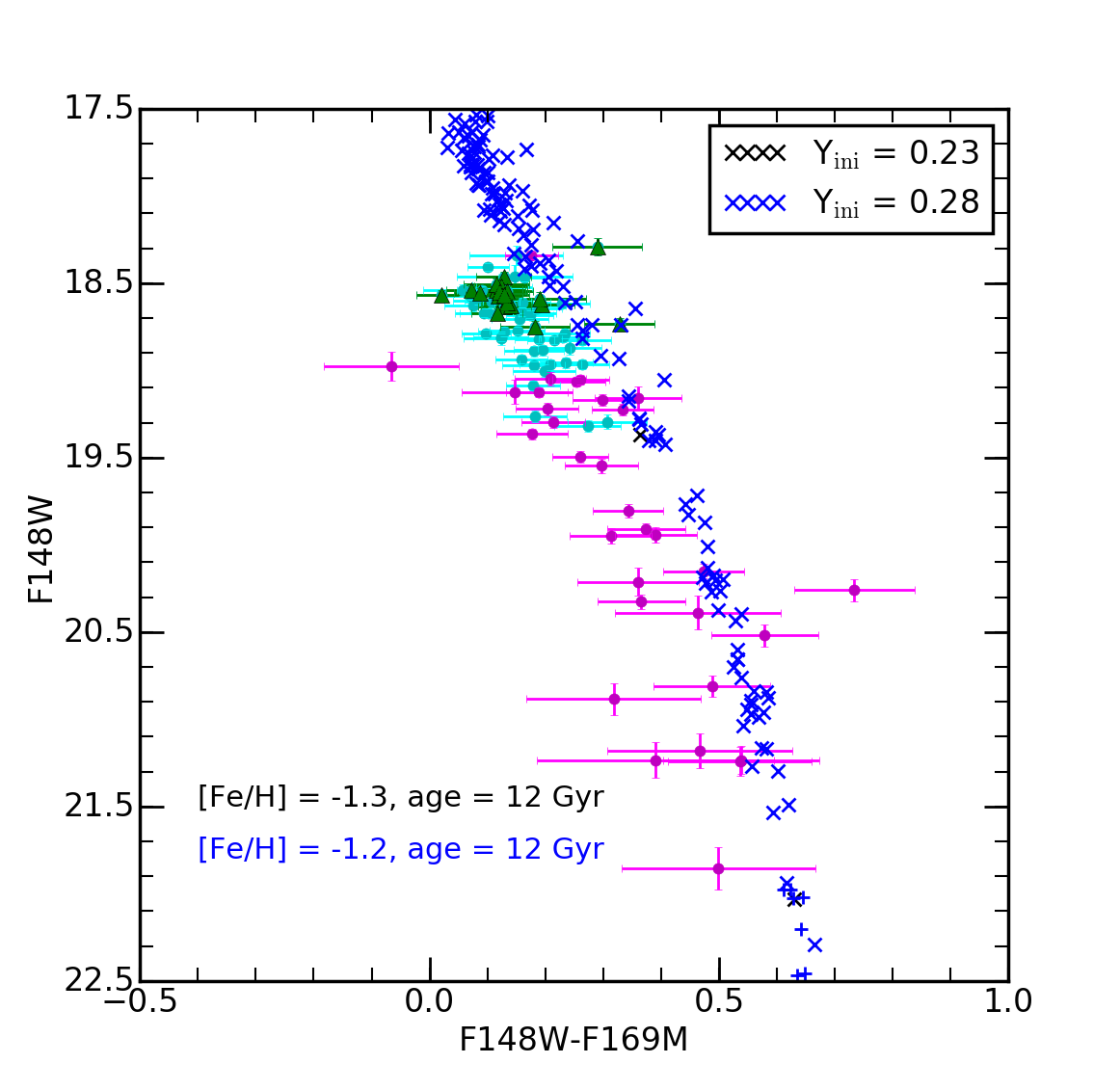}
\caption{FUV CMD based on UVIT photometry the F148W and F169M filters. BaSTI isochrones at three different ages are shown in the
top panel. The synthetic HB from the $Y^2$ models are shown in the bottom panel. }
\end{center}
\end{figure}

\section{Spatial Distribution of HB Stars}
\label{sec:spatial}

\cite{Milone2008} divided stars on the sub-giant branch into faint (fSGB) and bright (bSGB) populations, with
45\% and 55\% of the stars belonging to these two components. They found the HB to be bimodal as well, with 
63$\pm$7\% and 37$\pm$9\% associated with the red and blue branches, respectively. In our data, cross matched the HST data, we 
detect 232 HB stars in total, with 131 stars ($\sim$56\%) on the RHB and and 101 stars ($\sim$44\%) on the BHB. These fractions are similar to those found by 
\cite{Milone2008} for SGB and HB stars. The spatial distribution of the SGB and HB components were found
to be same by \cite{Milone2008}. The radial distribution of HB stars detected by UVIT and cross matched to HST are shown in Figure~14. The upper panel
shows the distribution of V-BHB, H-BHB and RHB, with the same color code as in the earlier figures. In the bottom panel, we have combined
the V-BHB and the H-BHB stars in to a single group (BHB) and their radial distribution (blue line) is shown along with the distribution of
RHB stars (red line). We note that the inner regions ($\le$ 1$^\prime$) have more BHB stars. Beyond $\ge 1\farcm4$, we see
an excess of RHB stars, as suggested by the cross-over of the two distributions. 
A K-S test for this pair of distributions (bottom panel) returns a probability of 2\%, which suggests that the two populations differ in their
radial distribution. The data used here for K-S test are those detected by UVIT from the original HST sample. Due to crowding, we were not
able to detect stars in the central $\sim10\arcsec$, highlighting that our data suffers from incompleteness in the core. \cite{Milone2008} found that the probability that
the BHB and RHB stars are drawn from the same sample is 19\%. Therefore, radially, the BHB and the RHB distributions may only be marginally different.

In Figure~15, we show the spatial distribution of the BHB stars in the upper panel and the histogram of the azimuthal distribution in the lower panel.
We converted the HB stars to an XY coordinate system with origin at the cluster center. In the upper panel, we have color coded the stars according
to their azimuthal angle, $\phi$, which increases towards the positive Y axis, starting from the negative X axis. 
The inner region with data incompleteness ($\sim 10\arcsec$ radius) is indicated. We have also indicated the half-light radius 
of the cluster as the dotted circle, with a radius of $r_h =$ 30\farcs6 \citep{Watkins2015}. The figure suggests that the BHB stars
show a non-uniform azimuthal distribution. In the lower panel, the histogram is shown to bring out the details, where we have show a bin width 
of 30$^o$, in azimuth angle. We have also shown the average number as the straight line, and the hashed area denotes the 1$\sigma$ width.
We see that two bins are below  and one bin is above this width. Therefore, there are less stars in the 60-90$^o$ and 150-180$^o$ bins, whereas there
are more stars in the 180-210$^o$ bin.
 
In Figure~16, we present a similar test for the RHB stars. This also
points to an azimuthally non-uniform distribution, albeit one that differs from the BHB stars. The spatial plot suggests that 
the RHB stars outside the half-light radius are preferentially located in three specific value of $\phi$. This is indicated by the histogram which shows 
a wavy pattern with three peaks and two troughs. Inside the half-light radius, the RHB stars are located mostly in the first and fourth quadrants, which is different 
from the distribution seen outside. The histogram shown suggests that two bins (120-150 and 240-270) have less stars, whereas the 150-180 bin has more stars
than 1$\sigma$ width. Thus, the RHB stars are also found to show a non-uniform azimuthal distribution. Both the BHB and RHB stars are 
azimuthally asymmetric, suggesting that they perhaps are not well mixed. 
The spatial distribution presented by \cite{Milone2008} shows similar features (i.e., see their Figure 7). The analysis provided here uses all stars outside the central 10\arcsec. The missing inner stars could alter the radial distribution, but may not affect
the azimuthal distribution. As the number involved is not large, we do not attempt to derive statistical significance of the azimuthal variation.
Our finding could provide additional support for the possibility that this cluster is a merger remnant \citep{Carretta2011}.

\begin{figure}
\begin{center}
\includegraphics[scale=0.6]{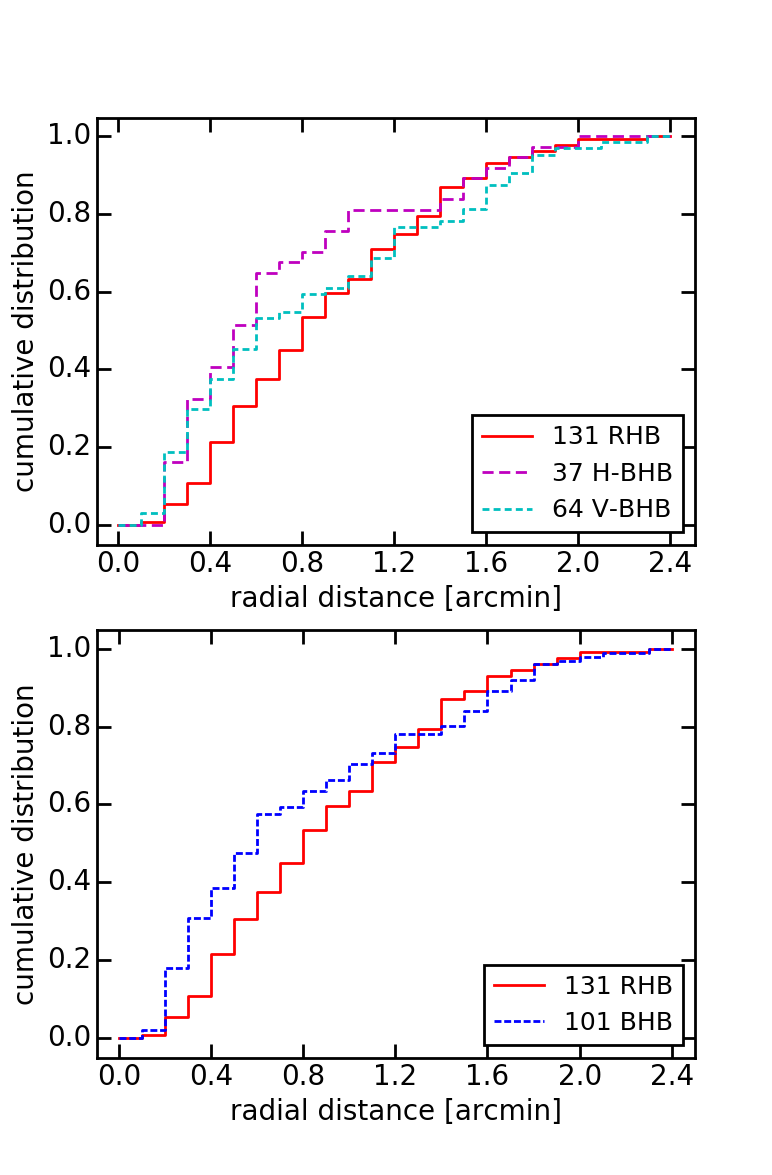}\\
\caption{The radial distribution of HB stars. The upper panel shows the distribution of V-BHB, H-BHB and RHB stars. In the lower
panel, the V-BHB and the H-BHB stars are combined into a BHB sample and shown along with their RHB counterparts.}
\end{center}
\end{figure}
\begin{figure}
\begin{center}
\includegraphics[scale=0.3]{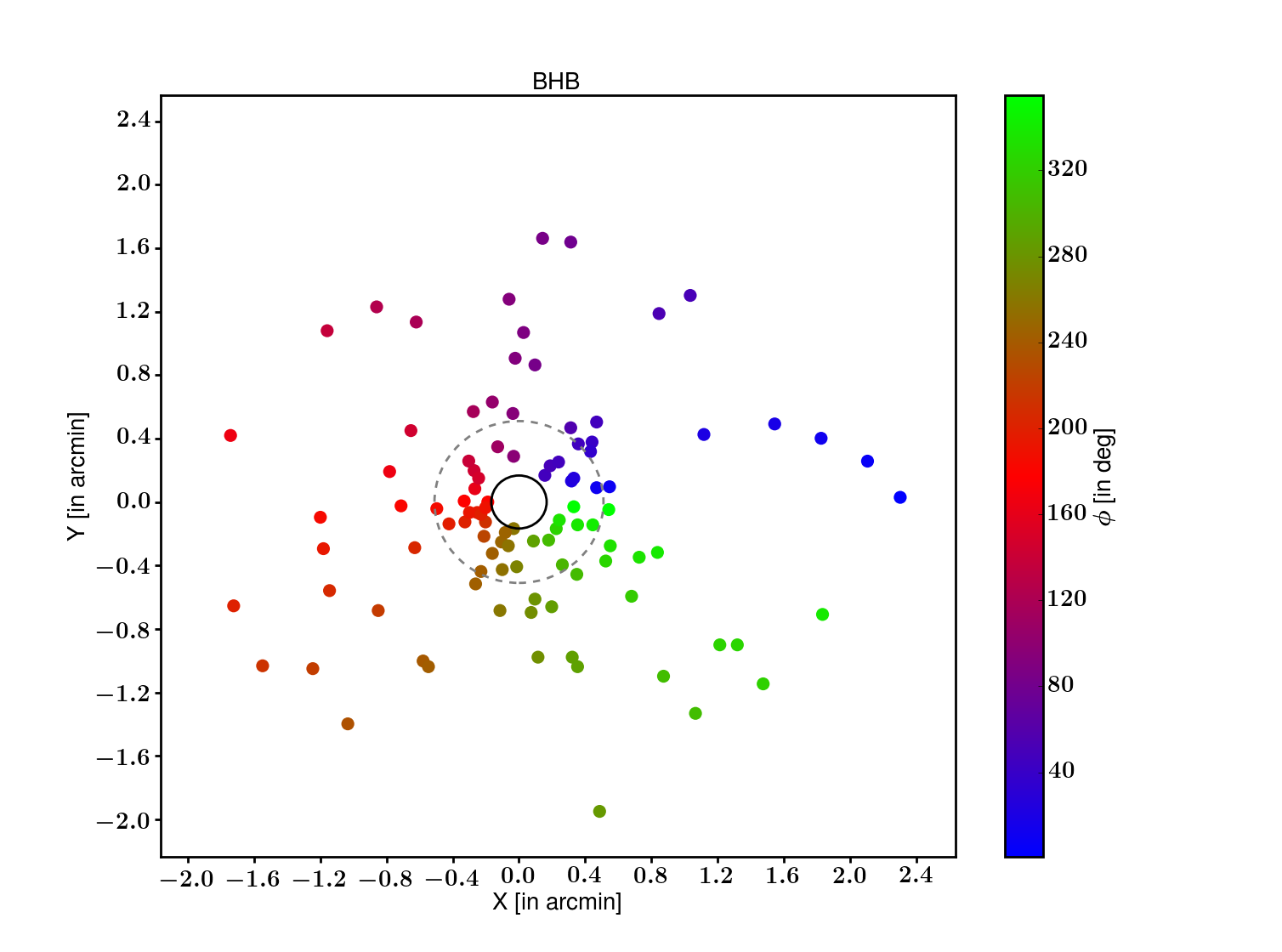}
\includegraphics[scale=0.34]{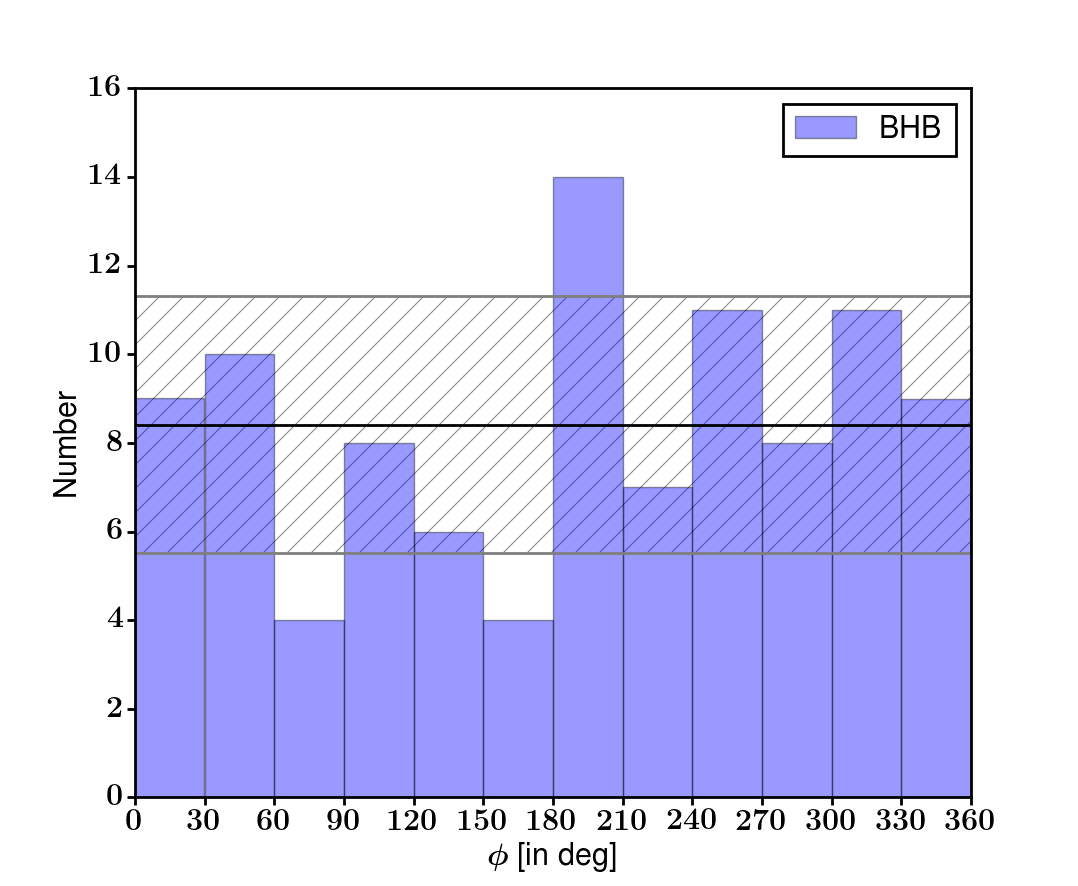}
\caption{The spatial distribution of BHB stars in NGC 1851. On the left, we plot the distribution of BHB star in the cluster. The origin marks the cluster center, 
while the inner circle (10\arcsec) corresponds to the region with significant sample incompleteness, and the
outer circle corresponds to the half-light radius (30.6\arcsec, \cite{Watkins2015}). The panel on the right shows the histogram of position angles, using bin width of 30$^\circ$. The straight line shows the mean value and the hatched region shows the 1$\sigma$ error width.}
\end{center}
\end{figure}
\begin{figure}
\begin{center}
\includegraphics[scale=0.3]{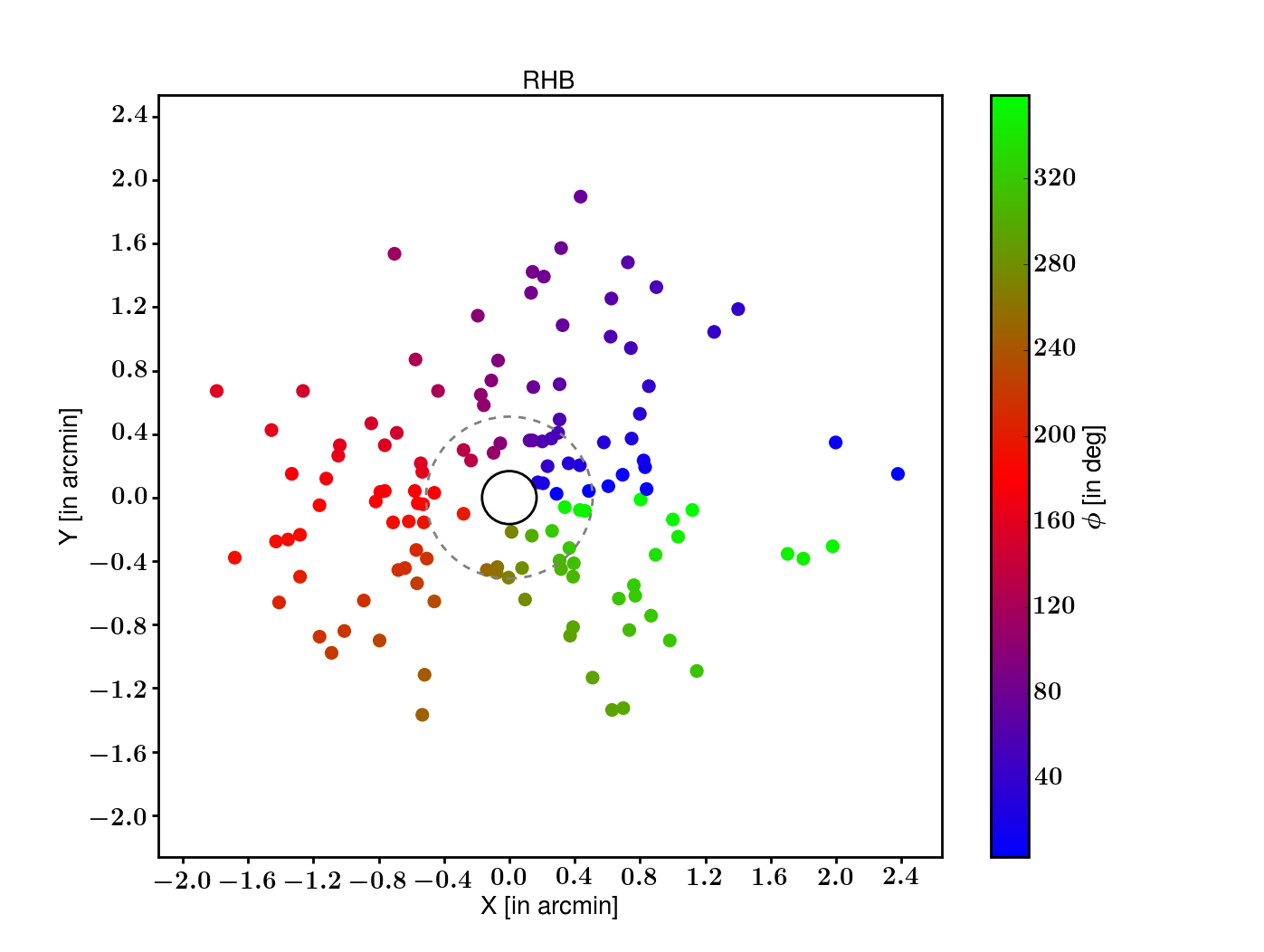}
\includegraphics[scale=0.34]{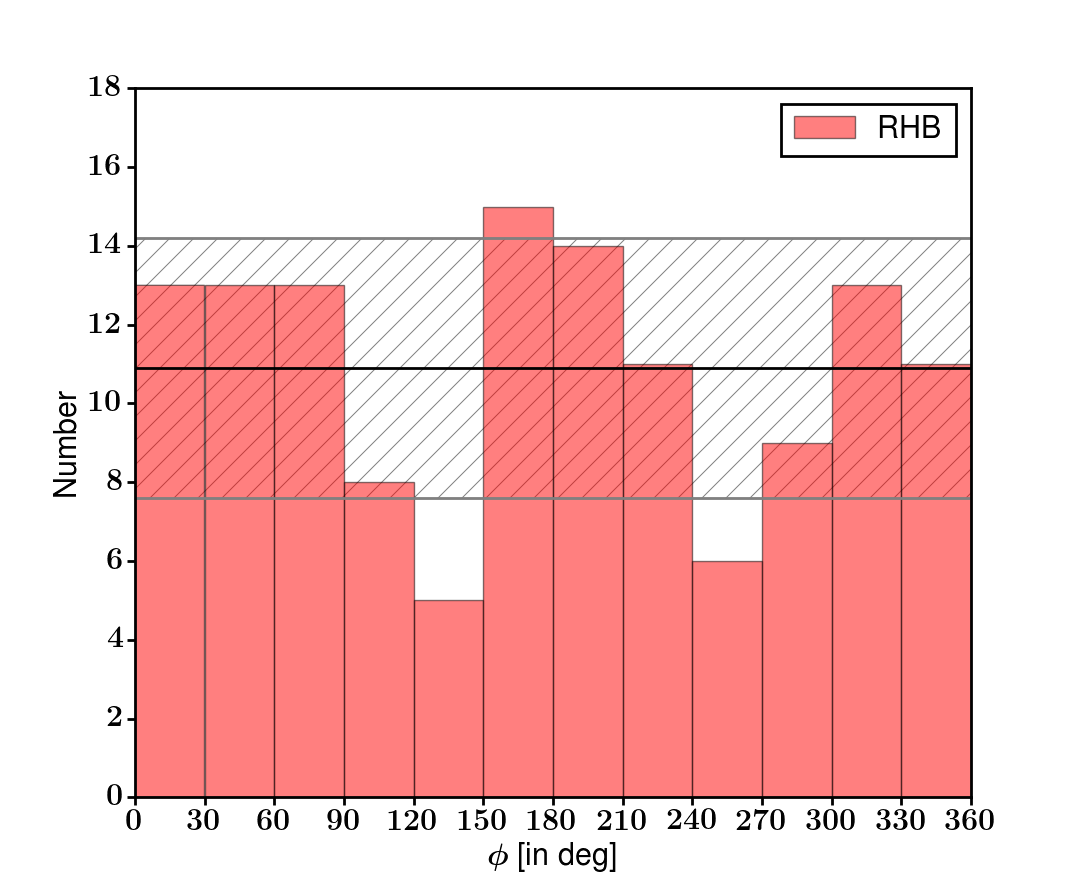}
\caption{Same as Figure 14, except showing the distribution of RHB stars within the cluster.}
\end{center}
\end{figure}

\begin{figure}
\begin{center}
\includegraphics[scale=0.55]{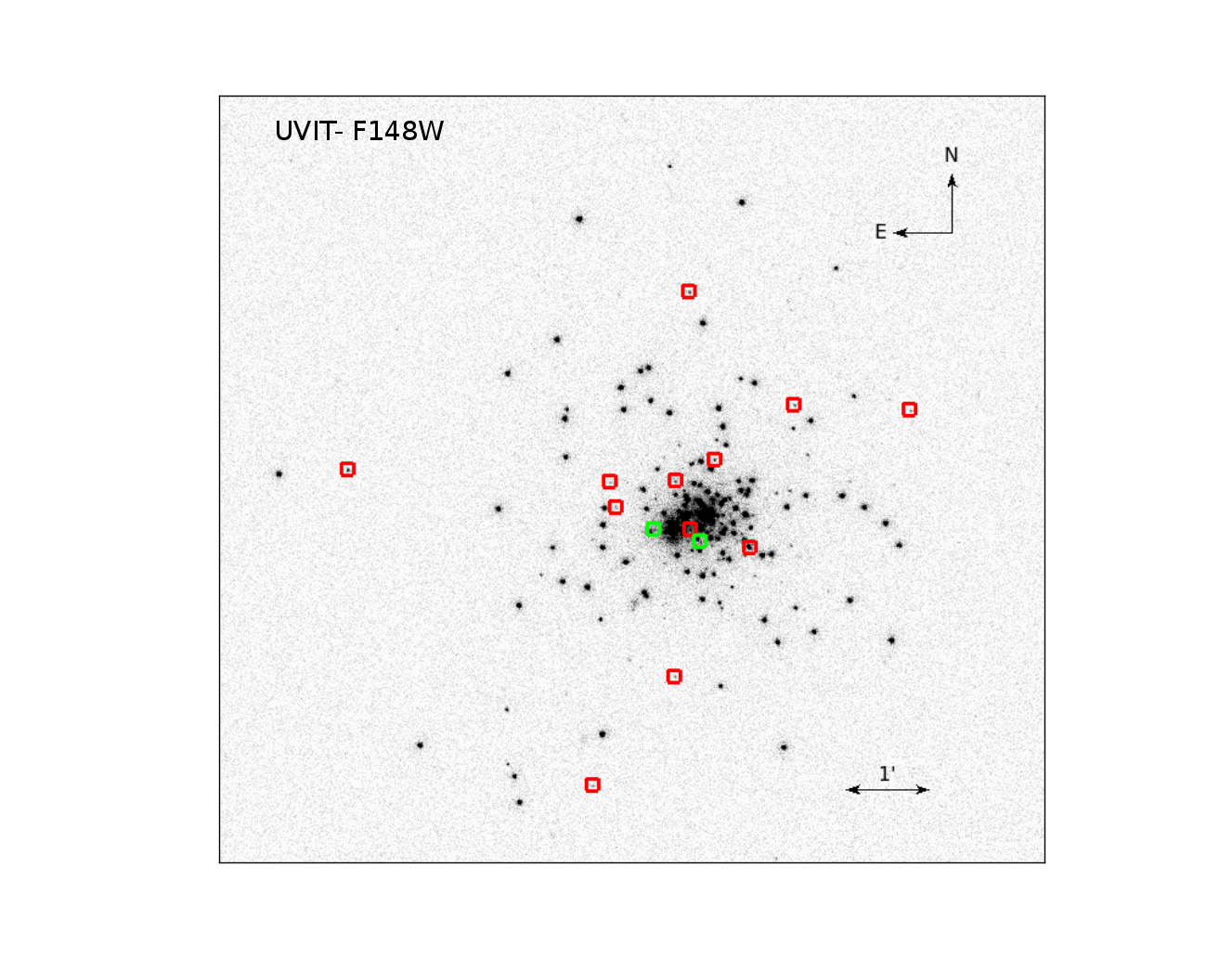}\\
\caption{RR Lyrae stars in NGC 1851 identified from our the FUV UVIT images. Two newly detected variables are shown in green. North is
up and east is to the left in this image, which measures 1\arcmin~on a side.}
\end{center}
\end{figure}
\begin{figure}
\begin{center}
\includegraphics[scale=0.55]{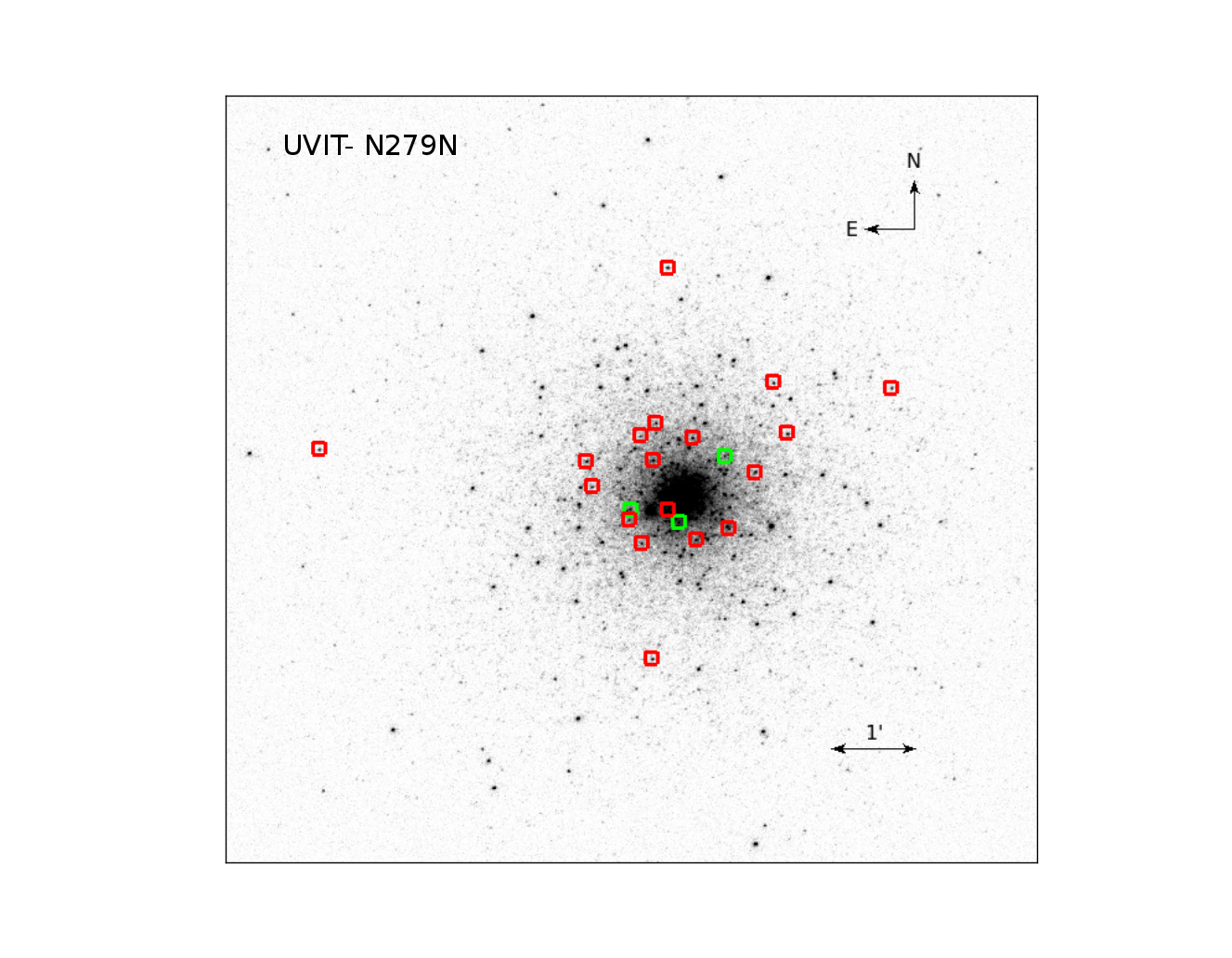}\\
\caption{RR Lyrae stars in NGC 1851 identified from our the NUV UVIT images. Three newly detected variables are shown in green. North is
up and east is to the left in this image, which measures 1\arcmin~on a side.}
\end{center}
\end{figure}
\begin{figure}
\begin{center}
\includegraphics[scale=0.4,angle=90]{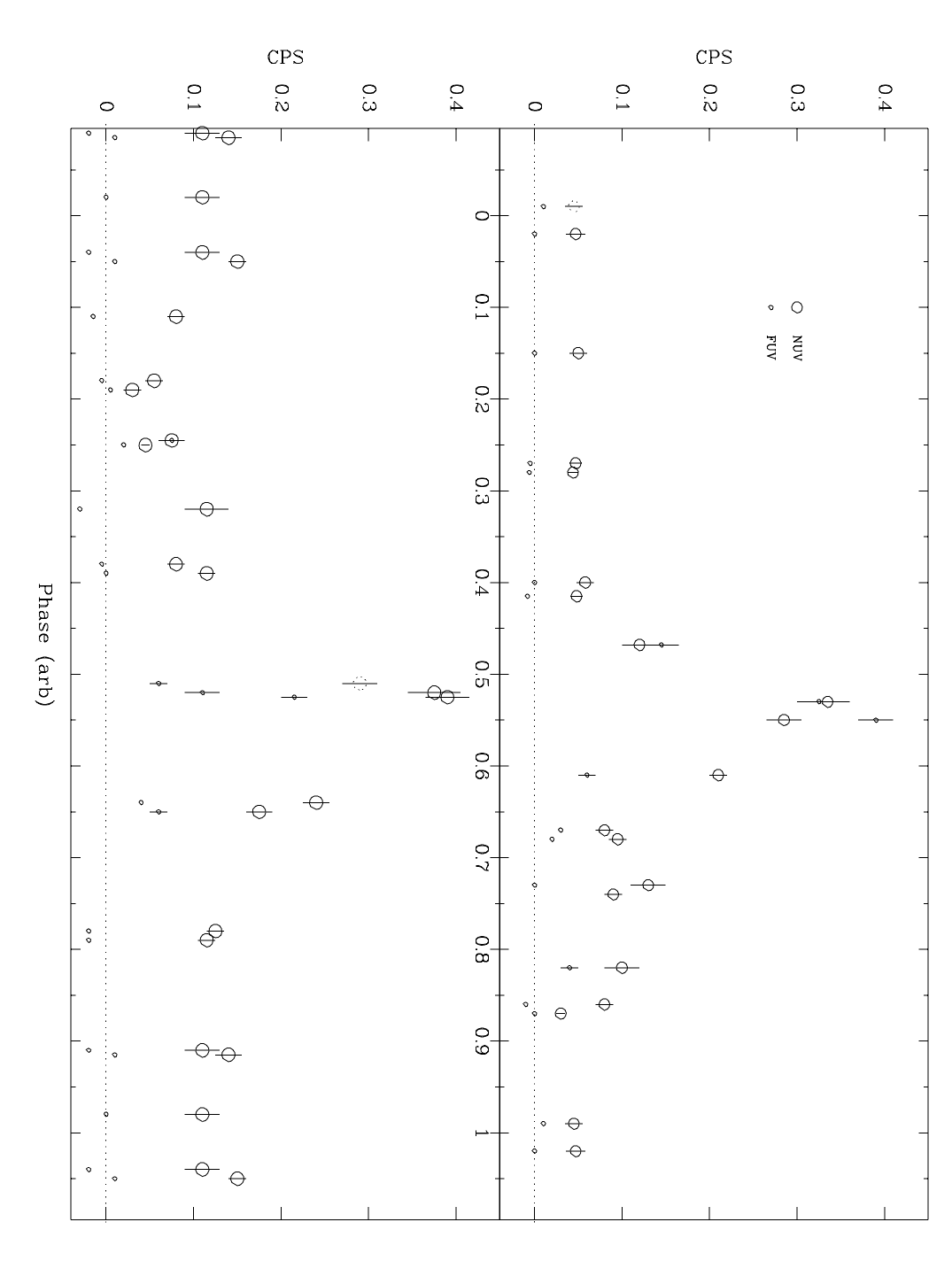}
\includegraphics[scale=0.4]{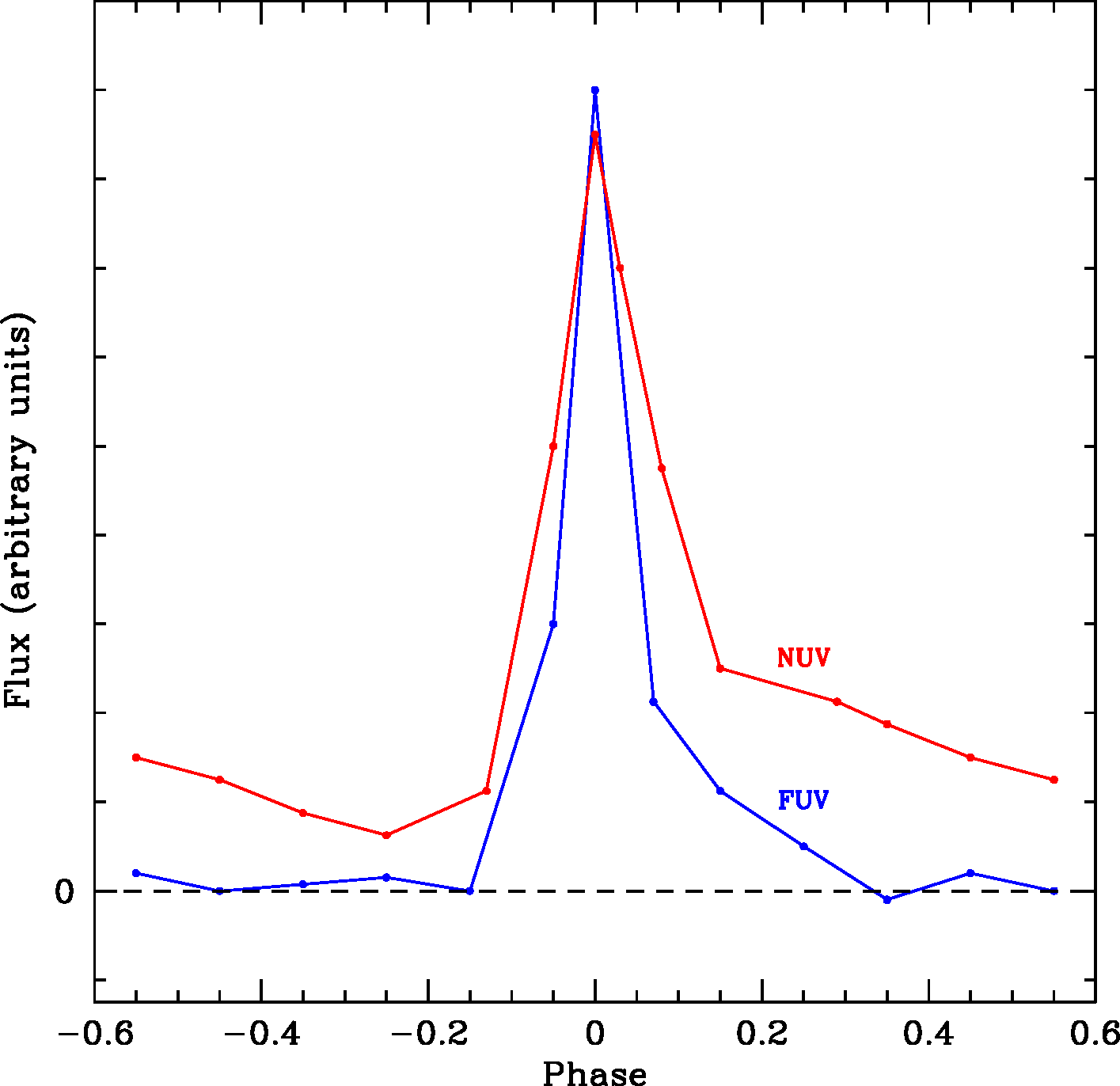}
\caption{Upper panel shows the full light curves of two RR Lyrae stars in NGC 1851. These full light curves and the rest of the partial light curves were used to create a typical
light curve for the FUV and NUV, as shown in the lower panel. }
\end{center}
\end{figure}

\section{UV Variability of RR Lyrae stars}
\label{sec:rr}

\cite{Oosterhoff1939} showed that Galactic globular clusters, according to their RR$_{ab}$ Lyrae stars, divide in two distinct 
groups --- one with RR$_{ab}$ periods of 0.55 day (Oosterhoff type I or OoI) and the other with periods of 0.65 day (Oosterhoff 
type II or  Oo II) period for RR$_{ab}$ stars. Oo I clusters have around 75\% of their RR Lyrae pulsating in the fundamental mode,
whereas for Oo II types, it is about 50\%. Oo I clusters tend to be more metal-rich and host fainter RR Lyrae variables than OoII 
clusters \citep{Caputo2000}.

In the specific case of NGC 1851, \cite{Walker1998} identified 29 RR Lyrae variables in the cluster and found the ratio of first-overtone RR Lyrae 
to fundamental mode RR Lyrae stars to be consistent with Oo II clusters (although near the end of Oo I distribution).  This cluster is known 
have the most extreme (long) period for RR$_{ab}$ stars, this is considered as one of the unusual OoI object \citep{Downes2004, 
Kunder2013}. It has been difficult to conclude whether the  differences in the intrinsic magnitudes of RR Lyrae variables in Oo I 
and Oo II clusters are caused by something other than just metallicity, such as age or helium content. In this cluster, where an internal spread 
in [Fe/H] is modest, at best, \cite{Kunder2013} suggested that a difference in helium
abundance among the RR Lyrae variables could explain the observed periods, amplitudes and brightnesses.
In the optical, RR Lyrae stars have typical amplitudes of 0.5--1.0 mag, whereas our UV data shows much larger variations. For instance,
\citep{Downes2004} identified 11  RR Lyrae-like variables in the core of the cluster using the HST FUV data. They found the
that the amplitudes of these stars to be as large as 4 magnitudes. Their Figure 3 show the light curves of RR Lyrae stars in the FUV
(although these are not complete light curves). 

Our UVIT observations of NGC 1851 were taken over multiple orbits as consequence of the early operations of Astrosat. Variable stars
were thus sampled over a range of their cycles that would not have been possible in a single pointing. 
We used the RR Lyrae stars from \cite{Walker1998} and \cite{Kunder2013} to cross match our sample. The RR Lyrae stars that we have
detected are shown in Figures 17 and 18. In all, we find 12 known and two new variables (red and green squares, respectively) in the 
FUV. In the NUV, we find 18 previously known and three new variable stars.  This brings the total of RR Lyrae like variables  in this cluster
to 43: i.e., 29 RR Lyrae stars were known from \cite{Walker1998} and another 11 from \cite{Downes2004}. We present full light curves
of two RR Lyrae stars in the FUV and NUV filters, in the upper panel of figure 19. These light curves demonstrate the structure of the variability in the FUV and the NUV.
It can be seen that , in the FUV, the RR Lyrae stars are almost non-detectable except during the maximum light,
whereas in the NUV, there is a measurable flux through the whole cycle.  We can also notice that a larger 
amplitude in the FUV, than that in the NUV. In order to bring out these features of the light curves, we created a 'typical' light curve profile.
The `typical' light curve is a mean of several individual ones created by a) cross correlating the phased plots to match the maxima, and 
b) taking the mean of the NUV and FUV phase-shifted values in bins of 0.05 phase. This is presented in the 
lower panel of figure 19.
The characteristic of the UV light curves is that the maxima are much sharper and the slow decline seen
in visible wavelengths is almost disappeared.
 

Our analysis of the HB population suggests that there could be RR Lyrae stars belonging to the RHB as well as the BHB stars. 
As noted above, our results suggest that there is an age and/or Helium difference between the BHB and the RHB. Our model HB population
shows the presence of two Y$_{ini}$ population in the region of the RR Lyrae stars.
Therefore, our analysis supports a difference in age or/and $Y_{ini}$ as the possible reason for the RR Lyrae
 properties in the cluster, and perhaps the Oosterhoff dichotomy.

\section{Discussion}
\label{sec:discussion}

In this study, we have presented the 
FUV and NUV CMDs which are obtained from the UVIT on ASTROSAT. The superior image quality of the UVIT, along with
the filter systems in the FUV and NUV channels have made this work possible. The UVIT images shown in the Figures 1, 2 and 3 demonstrate that the UVIT is
capable of producing excellent UV images of Galactic globular clusters. The calibration of the UVIT filter
systems are complete \citep{Tandon2017} and the effective areas as well as the zero-points are used to create the CMDs as well as the isochrones in the UVIT filters. 
We have also demonstrated that simultaneous observations in FUV and NUV
help in detecting variable stars. Our study has also added three RR Lyrae like stars to the previous catalog of 40 variables. 

As it is difficult to identify evolutionary sequences in the UV CMDs, we used the HST-ACS data to cross match and identify them. We detect mostly HB stars and 
a few blue straggler stars.  We divided the detected HB into the RHB, V-BHB and H-BHB stars. 
 We also detect a few blue hook stars in this cluster, appearing at the 
end of the vertical extension of the HB and the temperature is estimated to be $\sim$ 12,000K. 
This study demonstrates the ability of UV studies to bring out the details of HB population in globular clusters. 
We have analysed the UV CMDs with the help of isochrones generated in the UVIT filters, using the FSPS code \citep{Conroy2009}.
We compared the CMDs with Padova as well as BaSTI isochrones,  and the synthetic HB using the Helium enhanced $Y^2$ models. 

NGC 1851 is one of the many globular clusters now known to host multiple stellar populations. Evidence in this particular cluster comes 
primarily from the detection of two SGB sequences in the HST data \citep{Milone2008, Piotto2012}. Spectroscopic
studies performed on RGB \citep{Carretta2011} and HB stars \citep{Gratton2012} support the existence of multiple populations. 
We used a distance modulus of 15.52 mag \citep{Cassisi2008}. As the reddening is minimal towards this cluster \citep{Harris1996}, 
we have not incorporated reddening correction. The fit between the observed data and the models do have a  slight dependency on these parameters.
We identify and locate the full stretch of the HB stars in the UV CMDs, in the inner 4\farcm0 diameter of the cluster.
BaSTI models suggest that the  HB populations in NGC 1851 may have  an age difference of $\sim$ 2 Gyr, for a constant metallicity of
[Fe/H] $\sim$ $-$1.0~dex and Y$_{ini}$ $\sim$ 0.25. 
We are able to characterise two HB populations using the  $Y^2$ models, and the generated synthetic HB.
The best fits were obtained for an age range of $\sim$ 12 Gyr,  with a more or less
similar [Fe/H] of $\sim$ $-$1.3 - $-$1.2 and for two value of Y$_{ini}$,  $\sim$ 0.23 and 0.28.
The take away point is that both the models do not support the presence of super Helium rich stars in the HB. There  could be a moderate age difference ($\sim$ up to 2 Gyr) or  a moderate helium enrichment  ($\sim$ up to 0.05 dex)  among the HB stars. This is the first time, the HB parameters are estimated for this cluster from UV observations. These are in agreement
with those estimated from optical studies. 
Our observations are not deep enough to detect the sub-giant branch, which could have added more constraints on the estimated parameters. We plan to carry out deeper observations of more globular clusters,  as part of a broader program to carry our UVIT imaging of Galactic globular clusters.

The RHB and BHB stars, which were found to have a marginal difference in their radial distribution, were  found to have a non-uniform azimuthal 
distribution outside the cluster's central 10\arcsec. The uncorrelated and individually non-uniform distribution of 
the BHB and RHB stars suggest that these two populations are not very well mixed. This could provide support for the hypothesis that 
this cluster is a merger remnant.  We note that if the merger is recent, the populations could still remain not well mixed.The merging between
the two progenitor clusters could have  occurred recently, as the relaxation time at the half mass
radius is about 0.3 Gyr \citep{Harris1996}, whereas the full cluster relaxation takes about 2-3 relaxation times \citep{Carretta2011}. 

Since UVIT is operated in the photon counting mode, we were able to detect variability of known RR Lyrae stars in both the FUV and NUV bands. We are
able to create typical light curve in FUV and NUV wavelengths. These observations confirm the large amplitude variations of RR Lyrae stars previously 
seen in UV passbands, when compared to the optical. We find that the detected RR Lyrae stars may be belong to the RHB or the BHB.
As mentioned earlier, RR Lyrae stars in this cluster are known to belong to Oo I, with properties similar to Oo II. As we find that age and helium abundance to be
the main difference among the HB stars in this cluster, then it is likely that the differences among the RR Lyrae stars and the Oosterhoff classification may be due to these two parameters. 


\section{Conclusions}
\label{sec:conclusions}

The first globular cluster images from the UVIT on the Indian multi-wavelength astronomy observatory, ASTROSAT, is presented here.
The superior image quality of the telescope helps to almost resolve the core of the globular cluster NGC 1851. The large field of view
of the UVIT helps to detect the UV stars beyond the central region of the cluster. The PSF photometry
of $\sim$ 300 stars in two FUV filters and $\sim$ 2300 stars in one NUV filter are used to make UV CMDs. We are able to compare and match
the HST detected stars in the center of this cluster. Thus, we prove that the UVIT images in the FUV, in combination with the HST data,
is a very powerful tool to understand the stellar population in the core of globular clusters. We also generated isochrones in the 
UVIT filter system and are used to compare the observed CMDs, particularly the HB distribution. We have detected the full HB extent,
including the blue hook stars and RR Lyrae stars. We have identified 3 new RR Lyrae like variable in the cluster, which raises the
number of RR Lyrae like stars to 43. The simultaneous observation of the FUV and NUV channels in the photon counting mode helped in the identification of
 RR Lyrae stars in the FUV and the NUV bands. We used the Padova and BaSTI isochrones to compare the observed HB.
The BaSTI models suggest two populations among the HB stars, which are likely to have similar metallicity ([Fe/H] $\sim$ $-$1.0) and He abundance, 
but with two different ages (10 and 12 Gyr). The $Y^2$ models and the generated synthetic HB suggest one population with 
 an age of  12 Gyr,  Y$_{ini}$  =0.23 , and [Fe/H] $\sim$  $-$1.3 and another population of same age (12 Gyr), a similar Fe/H] $\sim$ $-$1.2, but with enhanced Y$_{ini}$  = 0.28.
  The distribution of BHB and the RHB stars were found to  suggest that the populations are not well mixed, which could lend support to
the idea that NGC 1851 is probably a recent merger remnant.
 
\acknowledgments
UVIT project is a result of collaboration between IIA, Bengaluru, IUCAA, Pune, TIFR, Mumbai, several centres of ISRO, and CSA. Indian Institutions and the Canadian Space Agency have contributed to the work presented in this paper. Several groups from ISAC (ISRO), Bengaluru, and IISU (ISRO), Trivandrum have contributed to the design, fabrication, and testing of the payload. The Mission Group (ISAC) and ISTRAC (ISAC) continue to provide support in making observations with, and reception and initial processing of the data.  We gratefully thank all the individuals involved in the various teams for providing their support to the project from the early stages of the design to launch and observations with it in the orbit. C.C. acknowledges support from the National Research Foundation of Korea to the Center for Galaxy Evolution Research (No.2017R1A5A1070354). We thank the referee for encouraging suggestions.
\bibliographystyle{apj}
\bibliography{ngc1851}
\end{document}